\documentclass[fleqn,usenatbib,useAMS]{mnras}

\usepackage{mathptmx}
\usepackage{threeparttable}

\usepackage[T1]{fontenc}
\usepackage{ae,aecompl}

\usepackage{graphicx}	
\usepackage{amsmath}	
\usepackage{amssymb}	
\usepackage{lscape}
\usepackage{multicol}
\usepackage{wasysym}

\newcommand{\msun}{$M_{\odot}$}

\newcommand{\kms}{km\,s$^{-1}$}

\newcommand{\vsini}{$v\,\,$sin$\,\,i$}

\newcommand{\nstars}{269}

\newcommand{\hei}{\ion{He}{i}}

\newcommand{\ci}{\ion{C}{i}}
\newcommand{\mgi}{\ion{Mg}{i}}
\newcommand{\mgii}{\ion{Mg}{ii}}
\newcommand{\baii}{\ion{Ba}{ii}}
\newcommand{\ali}{\ion{Al}{i}}
\newcommand{\sili}{\ion{Si}{i}}
\newcommand{\silii}{\ion{Si}{ii}}
\newcommand{\oi}{\ion{O}{i}}
\newcommand{\si}{\ion{S}{i}}

\newcommand{\caii}{\ion{Ca}{ii}}
\newcommand{\tiii}{\ion{Ti}{ii}}

\newcommand{\crii}{\ion{Cr}{ii}}

\newcommand{\mnii}{\ion{Mn}{ii}}
\newcommand{\ceiii}{\ion{Ce}{iii}}
\newcommand{\hgii}{\ion{Hg}{ii}}
\newcommand{\fei}{\ion{Fe}{i}}
\newcommand{\feii}{\ion{Fe}{ii}}
\newcommand{\srii}{\ion{Sr}{ii}}
\newcommand{\yii}{\ion{Y}{ii}}
\newcommand{\gaii}{\ion{Ga}{ii}}
\newcommand{\pii}{\ion{P}{ii}}
\newcommand{\ptii}{\ion{Pt}{ii}}
\newcommand{\zrii}{\ion{Zr}{ii}}
\newcommand{\xeii}{\ion{Xe}{ii}}
\newcommand{\priii}{\ion{Pr}{iii}}
\newcommand{\ndiii}{\ion{Nd}{iii}}

\title[HgMn stars]{The SDSS/APOGEE Catalog of HgMn Stars}

\author[S.~Drew Chojnowski et al.]{
S.~Drew Chojnowski,$^{1}$\thanks{E-mail: drewski@nmsu.edu}
Swetlana Hubrig,$^{2}$
Sten Hasselquist,$^{3,4}$
Rachael L. Beaton,$^{5}$
\newauthor Steven R. Majewski,$^{6}$
D. A. Garc\'ia-Hern\'andez,$^{7,8}$
and David DeColibus,$^{1}$
\\
$^{1}$Apache Point Observatory and New Mexico State University, P.O. Box 59, Sunspot, NM, 88349-0059, USA\\
$^{2}$Leibniz-Institut f\"ur Astrophysik Potsdam (AIP), An der Sternwarte~16, 14482~Potsdam, Germany\\
$^{3}$Department of Physics \& Astronomy, University of Utah, 115 1400 E, Salt Lake City, UT 84112, USA\\
$^{4}$NSF Astronomy and Astrophysics Postdoctoral Fellow\\
$^{5}$The Observatories of the Carnegie Institution for Science, 813 Santa Barbara Street, Pasadena, CA 91101, USA\\
$^{6}$Department of Astronomy, University of Virginia, P.O. Box 400325, Charlottesville, VA 22904-4325, USA\\
$^{7}$Instituto de Astrof\'isica de Canarias (IAC), E-38205 La Laguna, Tenerife, Spain\\
$^{8}$Universidad de La Laguna (ULL), Departamento de Astrof\'isica, E-38206 La Laguna, Tenerife, Spain\\
}

\date{Accepted XXX. Received YYY; in original form ZZZ}

\pubyear{2019}

\begin{document}
\label{firstpage}
\pagerange{\pageref{firstpage}--\pageref{lastpage}}
\maketitle

\begin{abstract}
We report on $H$-band spectra of chemically peculiar Mercury-Manganese (HgMn) stars obtained via the SDSS/APOGEE survey. As opposed to other varieties of chemically peculiar stars such as classical Ap/Bp stars, HgMn stars lack strong magnetic fields and are defined by extreme overabundances of Mn, Hg, and other heavy elements. A satisfactory explanation for the abundance patterns remains to be determined, but low rotational velocity is a requirement and involvement in binary/multiple systems may be as well. The APOGEE HgMn sample currently consists of {\nstars} stars that were identified among the telluric standard stars as those whose metallic absorption content is limited to or dominated by the $H$-band {\mnii} lines. Due to the fainter magnitudes probed by the APOGEE survey as compared to past studies, only 9/{\nstars} stars in the sample were previously known as HgMn types. The 260 newly-identified HgMn stars represents a more than doubling of the known sample. At least 32\% of the APOGEE sample are found to be binary or multiple systems, and from multi-epoch spectroscopy, we were able to determine orbital solutions for at least one component in 32 binaries. Many of the multi-lined systems include chemically peculiar companions, with noteworthy examples being the HgMn+Ap/Bp binary HD\,5429, the HgMn+HgMn binary HD\,298641, and the HgMn+Bp\,Mn+Am triple system HD\,231263. As a further peculiarity, roughly half of the sample produces narrow emission in the {\ci}~16895~{\AA} line, with widths and radial velocities that match those of the {\mnii} lines.

blank

\end{abstract}




\section{Introduction}
\label{sec:intro}

Whereas strong magnetic fields are the cause of the chemical peculiarities in classical Ap/Bp stars, the origin of the occasionally extreme heavy metal abundances in HgMn stars remains to be determined. The HgMn stars are typically defined by their optical spectra, in which overabundances of Hg \citep[up to 6 dex over Solar; e.g.,][]{white1976} and Mn \citep[up to 3 dex over Solar; e.g.,][]{smith1993} are the most obvious anomalies. Modern research shows that HgMn stars actually exhibit what appears to be a steady increase of overabundance with increasing atomic number \citep[][]{ghazaryan2016}. Other elements that can be readily confirmed as overabundant using ground-based data include P, Ga, Y, Xe, Pt, and Au. At the same time, light elements such as He are typically found to be underabundant \citep[e.g.,][]{castelli2004b}. Strong isotopic anomalies have been detected for the chemical elements Ca, Pt, and Hg, with the patterns changing from one star to the next \citep{hubrig1999,castelli2004b}.

Classification of HgMn stars based solely on the presence of {\hgii} and {\mnii} lines in the optical has naturally resulted in a very narrowly-defined and homogeneous sample compared to other chemically peculiar stars like the magnetic Ap/Bp stars and the (also poorly understood) metallic-lined Am/Fm stars. The HgMn stars cover a small spectral type range of B7--A0 main sequence stars \citep{renson2009}, and with an average rotational velocity of just $<${\vsini}$>\approx$30\,{\kms} (maximum {\vsini} around 100\,{\kms}), they are slow rotators as a rule.  Considering the HgMn star catalogs of \citet{schneider1981}, \citet{renson2009}, and \citet{ghazaryan2016}, as well as more recent finds by \citet{alecian2009,gonzalez2009,monier2015,monier2016,monier2017,caliskan2017,hummerich2018}, we estimate that there were 194 Galactic HgMn stars known prior to this work.

Over the past few decades it has become clear that the distributions of most or all elements on the surfaces of HgMn stars are inhomogeneous, i.e. these are stars with chemical/abundance spots. This was initially suspected by \citet{hubrig1995} based on the dramatically differing strength of the {\hgii}~3984\,{\AA} line for binary HgMn stars with similar spectral types but different inclination angles, and it was confirmed for the first time by \citet{malanushenko1996}, who found that the radial velocities of the {\hei} lines of $\alpha$\,And were variable with a period corresponding to the rotation period of the star. Recent high-resolution spectroscopic studies have subsequently shown that abundance spots on HgMn stars are the rule rather than the exception \citep[][etc.]{briquet2010,hubrig2010,makaganiuk2011,makaganiuk2012}. The spot patterns not only differ strongly from element to element, they are also highly variable, changing over timescales as short as a few months \citep[e.g.,][]{korhonen2013}. It is well known that in the case of Ap/Bp stars, the spot patterns are correlated with the topology of their strong, global magnetic fields. The existence of abundance spots on the comparatively non-magnetic HgMn stars is therefore quite perplexing. However, the problem may be partially resolved by the recent detection of variable, weak, tangled magnetic fields in two HgMn stars \citep{hubrig2020}.

Another key aspect of HgMn stars seems to be their high rate of binarity/multiplicity. Based on a check of literature radial velocity and binary orbit studies, \citet{hubrig1995} found that the multiplicity fraction may be as high as 67\% and that there was a preference for orbital periods shorter than 20 days. \citet{scholler2010} subsequently used high-contrast imaging to detect visual companions around 28 of 56 observed HgMn stars, suggesting a multiplicity fraction of at least 50\%. Recently, \citet{kervella2019} studied the multiplicity fraction of nearby stars by combining proper motions from the Hipparcos and Gaia catalogs to check for proper motion anomalies (PMa) indicative of one or more companion objects. Of the 155 HgMn stars included in the study, 60/155 are known visual multiple stars and 36 of these 60 exhibit significant PMa. Another 43 HgMn stars for which companions have not been visually identified also exhibit signficant PMa, leading to a total of 103/155 (66\%) definite or possible multiples, consistent with the findings of \citet{hubrig1995}.

The companions of HgMn stars are frequently also found to be chemically peculiar. The bright HgMn binary 46\,Dra is one of the more remarkable examples, since the companion is a nearly equal mass star that also shows the HgMn pecularities \citep{tsymbal1998}. The 41\,Eri, HR\,7694, and HD\,33647 systems are further examples of HgMn+HgMn binaries. A few systems also include magnetic chemically peculiar companions, with fields having been measured in the Am companion of $\chi$\,Lup \citep{mathys1995} and in the Ap companion of HD\,161701 \citep{hubrig2014}. The latter system is noteworthy in terms of being the only known example of an HgMn+Ap binary \citep{gonzalez2014}. More typically, the companions are found to be Am stars, especially the cooler ones \citep{ryabchikova1998}. 

In the HgMn binaries  with synchronously rotating components, the stellar surfaces facing the companion star usually display low-abundance element spots, whereas the surface of the opposite hemisphere, as a rule, is covered by high-abundance element spots \citep{hubrig2012}. Membership in binary and multiple systems may therefore be an important clue toward developing a better understanding of HgMn stars.

The puzzling pattern of chemical abundances and variability, as well as the poorly understood role of magnetic fields in the HgMn stars challenges our very understanding of the nature of these objects. In the following, we report on the detection of {\nstars} HgMn stars (only 9 previously known) in the $H$-band spectra obtained within the framework of the SDSS/APOGEE survey. This new sample represents a more than doubling of the number of known HgMn stars, and is generally an extension of the known sample to fainter magnitudes than have been probed in past studies. In addition to analyzing the $H$-band line content and line profiles of HgMn stars for the first time, we use the multi-epoch spectra available for most of the stars to measure radial velocities (RVs) and search for binaries. Where possible, we also estimate orbital parameters for the numerous single-lined (SB1), double-lined (SB2), and triple-lined (SB3) spectroscopic binaries among the sample.  

\section{Data} \label{data}

\subsection{SDSS/APOGEE $H$-band Spectroscopy}
The APOGEE instruments are 300-fiber, $R=22,500$, $H$-band spectrographs that operate on the Sloan 2.5m telescope \citep{gunn2006} at Apache Point Observatory (APO, APOGEE-N) and on the Du Pont 2.5m telescope \citep{bowen1973} at Las Campanas Observatory (LCO, APOGEE-S). Each APOGEE instrument records most of the $H$-band (15145--16960\,{\AA}; vacuum wavelengths used in this paper) on three detectors, with coverage gaps between 15800--15860\,{\AA} and 16430--16480\,{\AA} \citep{wilson2019} and with each fiber subtending a $\sim$2\arcsec diameter on-sky field of view. The typical exposure time for an APOGEE `field' (1.5$^{\circ}$--3$^{\circ}$ diameter area of sky) is one hour, or roughly the time needed to reach signal-to-noise ratios (S/N) of 100 for stars with $H\leq11$ mag. The fields are usually observed multiple times on different nights, months, or years, to accumulate signal for fainter targets, and the individual spectra are later combined to produce high quality spectra for chemical abundance analysis. \citet{majewski2017} provided a detailed overview of the APOGEE survey, and \citet{nidever2015} described the data reduction process. This paper makes use of spectra obtained during both SDSS-III \citep{eisenstein2011} and SDSS-IV \citep{blanton2017} and reduced by the 15$^{\rm th}$ SDSS data release \citep[DR15;][]{aquado2019} version of the reduction pipeline. 

The APOGEE targeting strategy was summarized by \citet{zasowski2013,zasowski2017}, but some of the relevant details will be repeated here. In each APOGEE observation, 15--35 telluric standard stars ($TSS$) are targeted to facilitate removal of telluric absorption features from the science spectra. The $TSS$ are selected quasi-randomly based on raw (un-dereddened) Two Micron All-Sky Survey \citep[2MASS;][]{skrutskie2006} $J-K$ color within an $H$-band magnitude range of $7.0<H<11.0$. There are also spatial restrictions to account for variability of the telluric absorption over the 2$^{\circ}$--3$^{\circ}$ fields, and during SDSS-III, they were often excluded if positioned within the projected radius of an open cluster since this might induce collisions with the fiber casings of cluster science targets. In general however, the $TSS$ are the combined bluest and brightest available stars in the field of view. As of July 8, 2019, a total of 33,838 $TSS$ had been observed 192,545 times for an average of 5.7 spectra per star. The HgMn sample described here is culled primarily from the $TSS$ sample, though with a few stars having been targeted as confirmed or candidate open cluster members during SDSS-IV. The HgMn sample at hand averages S/N=184 per individual spectrum, such that we are firmly operating in the high-S/N regime.

\begin{figure}
\includegraphics[width=1.0\columnwidth]{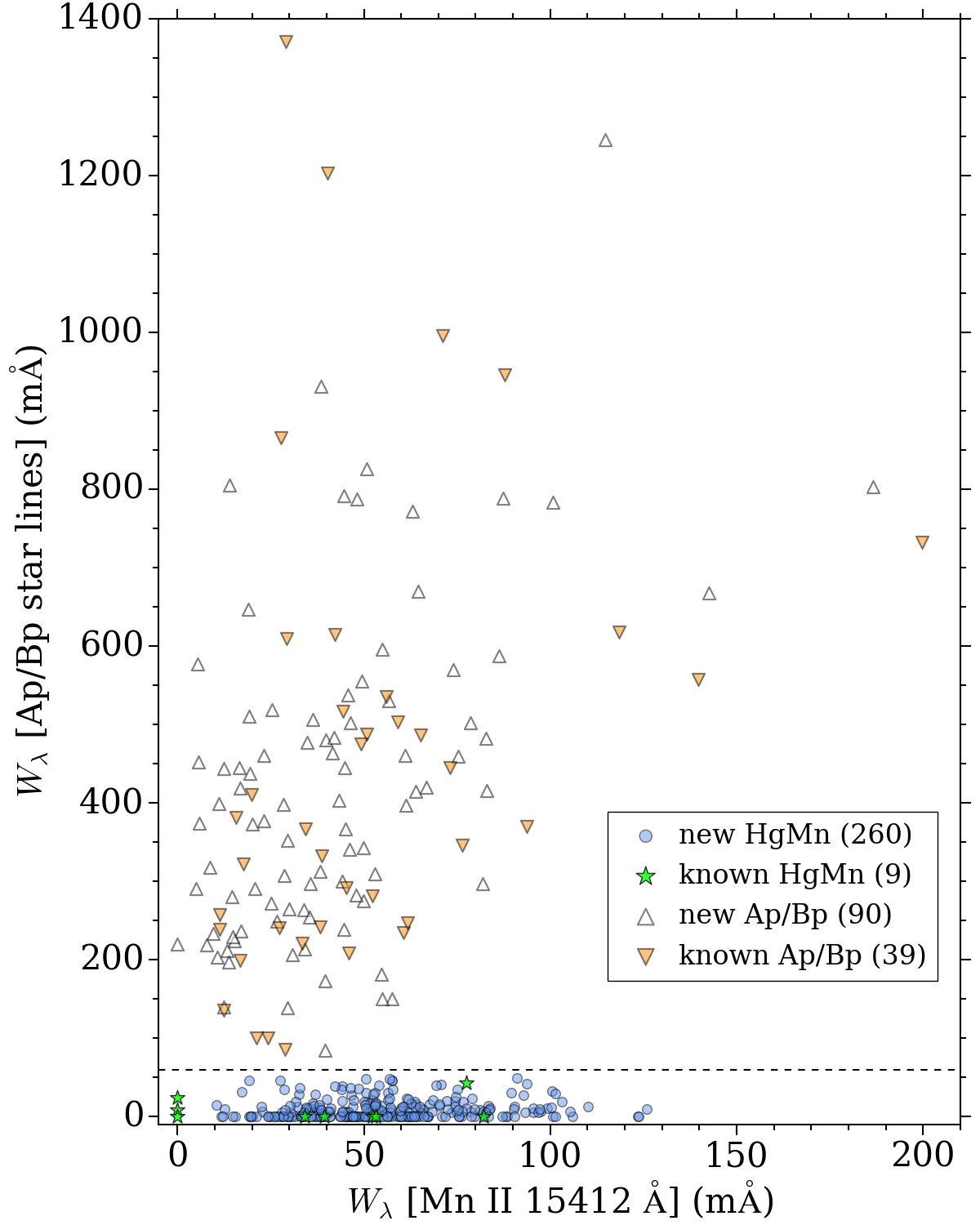}
\caption{HgMn versus Ap/Bp classification, based on comparison of the strength of lines that typically appear in Ap/Bp star spectra to the strength of {\mnii}~15413\,{\AA}. The ordinate is the sum of the equivalent widths of {\ceiii}~15961, 15965, {\crii}~15370, 15470\,{\AA}, and the unidentified lines at 15231 and 16184\,{\AA} \citep[discussed in][]{choj2019}. We classify as HgMn any stars below the dashed line at $y=60$\,m{\AA}. }
\label{eqwplot}
\end{figure}

\subsection{APO 3.5m + ARCES Optical Spectroscopy}
We also obtained optical spectra of 32 newly discovered HgMn stars, 15 previously known HgMn stars, and Vega using the Astrophysical Research Consortium Echelle Spectrograph \citep[ARCES;][]{wang2003} on the ARC 3.5m telescope at APO. In each exposure, ARCES covers the full optical spectrum (3500--10400 {\AA}) at a resolution of $R$=31,500. We used standard Image Reduction and Analysis Facility (IRAF\footnote{IRAF is distributed by the National Optical Astronomy Observatories, which are operated by the Association of Universities for Research in Astronomy, Inc., under cooperative agreement with the National Science Foundation.}) echelle data reduction techniques, including 2D to 1D extraction, bias subtraction, scattered light and cosmic ray removal, flat-field correction, wavelength calibration via Thorium--Argon lamp exposures, as well as continuum normalization and merging of orders. The majority of the new HgMn stars were observed on multiple epochs in order to improve the derived orbital solutions. Exposure times were tailored to ensure S/N$>$100 at 4000\,{\AA} in each observation. 

\begin{figure*}
\centering 
\includegraphics[width=1.0\textwidth]{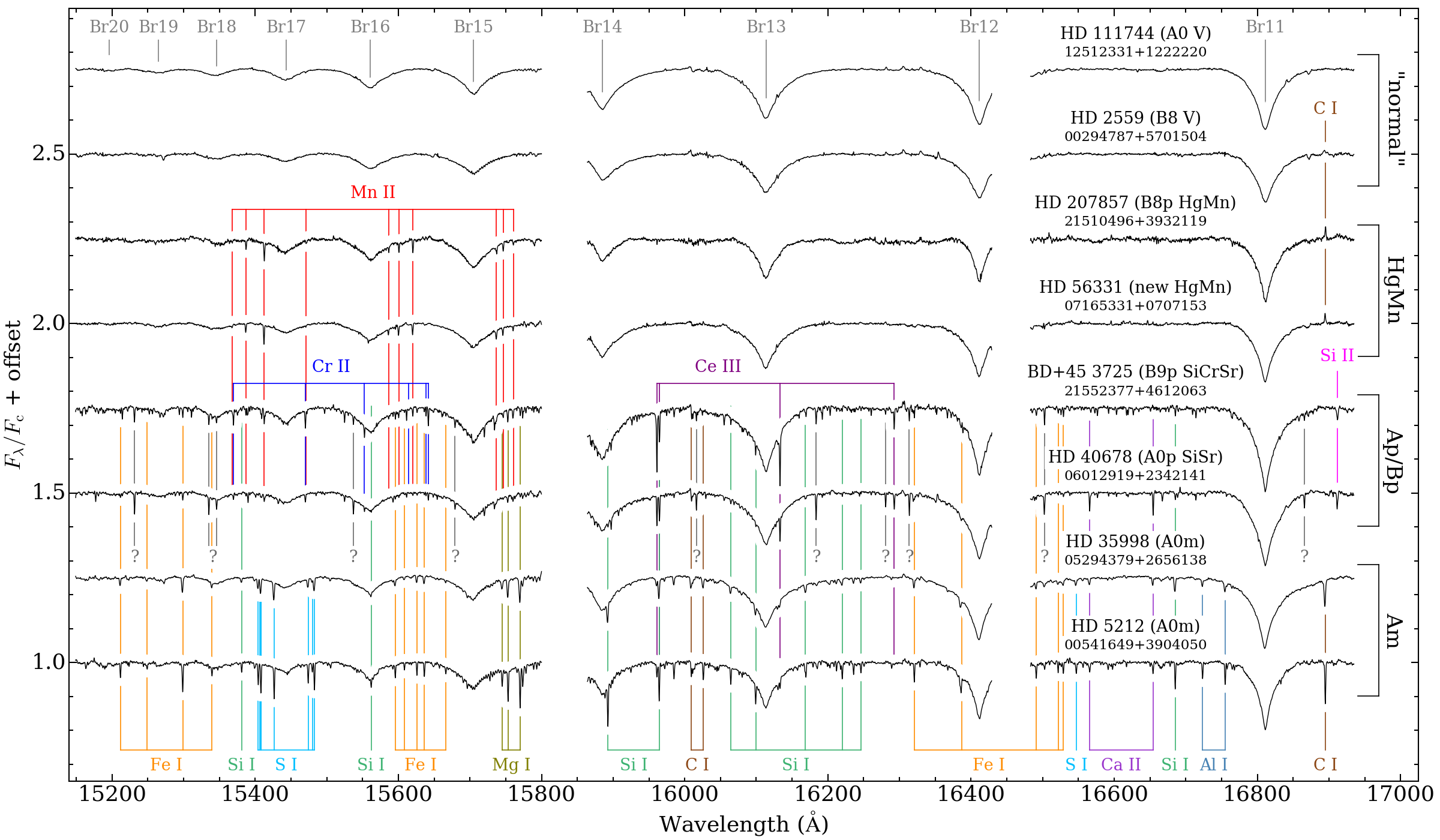}
\caption{Comparison of the $H$-band spectra of new and previously-known HgMn stars (HD\,207857 and HD\,56331) to presumably normal stars (HD\,2559 and HD\,111744), Ap/Bp stars (BD+45\,3725 and HD\,40678), and Am stars (HD\,35998 and HD\,5212). Literature spectral types and 2MASS designations are given at right, and the majority of strong spectral features have been labeled. The line content of the normal stars is mostly limited to the hydrogen Brackett series lines, though HD\,2559 exhibits weak emission in the {\ci}~16895\,{\AA} line. The HgMn stars have strong {\mnii} and emission in {\ci}~16895\,{\AA}. The Ap/Bp stars also have {\mnii}, but the strongest lines are from {\ceiii}, {\crii}, the unknown lines (labeled with question marks), {\silii}, and {\caii} (in the case of HD\,40678). The Am stars also exhibit {\ceiii}, but the {\mnii}, {\crii}, and unknown lines are never present. Instead, the `normal star lines' ({\ci}, {\sili}, {\si}, {\mgi}, {\ali}, and {\fei}) are overly strong.}
\label{specmontage}
\end{figure*}

\section{Sample Selection} \label{sample}
The HgMn star sample discussed in this paper was discovered among a much larger catalog of APOGEE Chemically Peculiar (ACP) stars that currently consists of $>2100$ stars collectively with $\sim$12,800 individual spectra. Assembly of the ACP catalog has been based primarily on detection of the {\ceiii} lines at 15961, 15965, and 16133~{\AA} (with weaker lines occasionally present at 15720 and 16293\,{\AA}), as described in \citet{choj2019}. These are often the strongest metallic features in the $H$-band spectra of Ap/Bp stars, and they are often the only metallic features present in the spectra of these stars. Roughly 26\% of the ACP sample consists of stars that do not exhibit the {\ceiii} lines and were instead added to the catalog due to prior classifications of chemical peculiarity in the literature. 

The {\ceiii} lines are also useful for identifying Am/Fm stars despite the `normal star lines' ($NSL$), i.e. the {\mgi}, {\sili}, {\fei}, {\ci}, {\si}, and {\ali} lines that are present in the APOGEE spectra of nearly all stars with effective temperature less than $\sim$8000\,K, always being stronger than {\ceiii} for Am/Fm stars. Both Ap/Bp and Am/Fm stars can exhibit the $NSL$ however, such that there is some ambiguity. Based on inspection of the spectra of the many hundred stars in the ACP catalog with literature Ap/Bp or Am/Fm classifications, the two groups can be distinguished based on a quite simple criterion. Namely, the ratio of {\ceiii}~15961\,{\AA} over the neighboring blend of {\ceiii}~15964.928\,{\AA} + {\sili}~15964.424\,{\AA} is always less than unity for stars classified as Am/Fm, and vice versa for those classified as Ap/Bp stars. Another equally viable classification method is to check for the presence of lines that are only detected in the spectra of Ap/Bp-classified stars. Good options for this are the {\crii} and {\mnii} lines, with perhaps the most useful option being the set of unidentified lines ($UL$) that are often prominent in Ap/Bp star spectra and that were used for magnetic field modulus measurement in \citet{choj2019}. Although we suspect the $UL$ are formed by one or more rare earth elements, the associated atomic data do not exist or are not publicly available. Regardless, these lines do not appear in the spectra of Am/Fm stars. 

One of the most severe outliers among the apparent Ap/Bp stars identified via {\ceiii} detection was the well known HgMn star HD\,207857, for which the metal absorption line content is limited to {\mnii} and {\ceiii}, but with the strengths of the {\mnii} lines more than doubling those of the {\ceiii} lines. This revelation prompted us to search the $\sim$40,000 APOGEE telluric standard stars for additional objects with {\mnii} absorption lines. This was done by measuring equivalent widths via direct integration of the flux contained in fixed 3\,{\AA} windows centered on the two strongest {\mnii} lines covered (15413\,{\AA} and 15620\,{\AA}). These lines conveniently fall in portions of the spectra that are unaffected by strong airglow emission or telluric absorption features. They are also well separated from any lines that might be expected to appear in the spectra of F-type or hotter stars, such that there is minimal confusion due to blending. 

\begin{figure*}
\centering 
\includegraphics[width=1.0\textwidth]{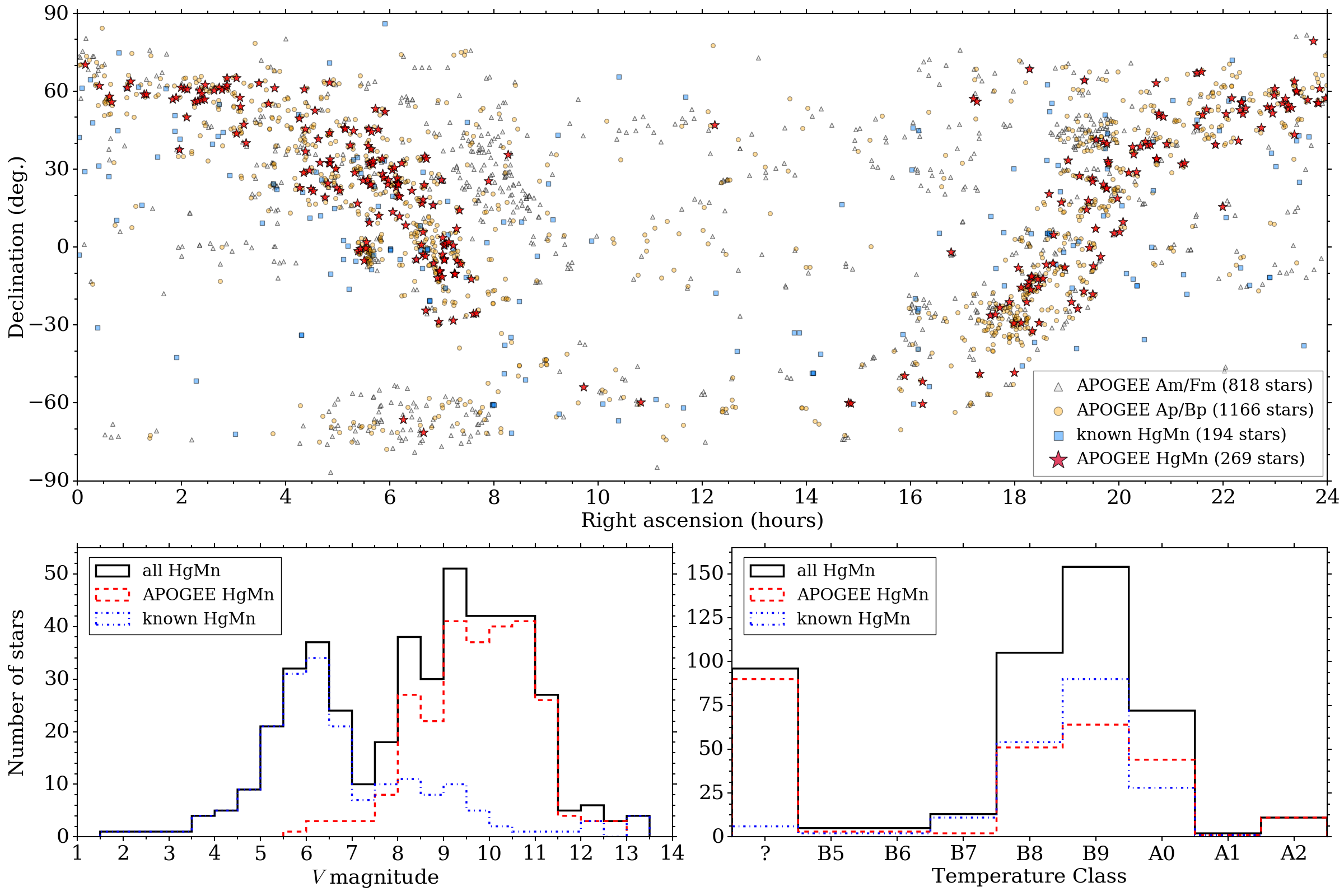}
\caption{(\emph{Top:}) Spatial distributions of the APOGEE chemically peculiar stars and the previously known HgMn stars. (\emph{Lower left:}) Histogram of $V$ magnitude for the HgMn stars. (\emph{Lower right:}) Histogram of literature spectral types for the HgMn stars. The column labeled ``?'' in the lower right panel indicates the almost 100 HgMn stars that lack a literature spectral type. }
\label{radecplot}
\end{figure*} 

\begin{table*}
\caption{Summary of the APOGEE HgMn sample. Only the first 10 rows are shown (full version available online).}
\centering
\begin{threeparttable}
\scriptsize
\label{startable}
\begin{tabular}{clcccccccrrll}
\noalign{\smallskip}\hline\hline
            &       &           &           &                             &           & $<$$W_{\lambda}$$>$ & $<$$W_{\lambda}$$>$ & $<$$W_{\lambda}$$>$ &          &                   &        &            \\ 
            &       &           &           &                             &           & {\mnii}             & {\ceiii}            & {\ci}               &          &                   &        & Literature \\ 
2MASS       & Other & 2MASS $H$ & $<$S/N$>$ & $N_{\rm RV}/N_{\rm obs.}$ & {\vsini} & 15413               & 15961               & 16895               & $<$RV$>$ & RV$_{\rm scat.}$ & Binary & Spectral \\ 
Designation & Name  & (mag)     &           &                             & ({\kms})  & (m{\AA})            & (m{\AA})            & (m{\AA})            & ({\kms}) &   ({\kms})        & Type   & Type     \\ 
\noalign{\smallskip}\hline\hline
00091817+7022314 & ... & \phantom{1}8.953 & 216 & 12/12 & 17 $\pm$\phantom{0}8\phantom{*} & \phantom{1}35 & ... & -33 & -19.1 $\pm$\phantom{0}2.1 & 4.7 & ... & ... \\ 
00245787+6212255 & ... & 10.479 & 112 & 11/12 & \phantom{0}4 $\pm$\phantom{0}4\phantom{*} & \phantom{1}23 & ... & -37 & -29.5 $\pm$\phantom{0}2.8 & 7.0 & SB1 & ... \\ 
00364423+5810121 & ... & 10.345 & 109 & 8/13 & 62 $\pm$\phantom{0}2* & \phantom{1}61 & ... & ... & -31.5 $\pm$\phantom{0}7.5 & 21.9 & ... & ... \\ 
00390709+5558188 & ... & 10.472 & 109 & 6/6 & 36 $\pm$\phantom{0}7\phantom{*} & \phantom{1}54 & ... & ... & -21.5 $\pm$\phantom{0}5.4 & 2.6 & ... & ... \\ 
00570626+6134007 & HD 5429 & \phantom{1}8.403 & 268 & 11/11 & 20 $\pm$\phantom{0}2* & \phantom{1}30 & ... & -47 & -3.2 $\pm$\phantom{0}2.2 & 62.5 & SB2* & B8\,III\tnote{1} \\ 
01011057+6355294 & Hilt 91 & \phantom{1}9.860 & 140 & 3/3 & 10 $\pm$\phantom{0}4\phantom{*} & \phantom{1}82 & ... & -30 & -27.7 $\pm$\phantom{0}1.6 & 0.1 & ... & B9\,III:\tnote{1} \\ 
01165086+5902504 & ... & \phantom{1}9.709 & 129 & 4/4 & 58 $\pm$\phantom{0}5\phantom{*} & \phantom{1}62 & ... & ... & -15.5 $\pm$\phantom{0}7.4 & 7.8 & ... & B8\tnote{1} \\ 
01191843+5902478 & HD 7844 & \phantom{1}8.317 & 261 & 6/6 & 39 $\pm$\phantom{0}2* & \phantom{1}79 & ... & -72 & -29.1 $\pm$\phantom{0}2.0 & 4.3 & ... & B8\tnote{1} \\ 
01494647+5653549 & ... & 10.600 & \phantom{5}88 & 3/3 & 15 $\pm$\phantom{0}5\phantom{*} & \phantom{1}33 & ... & ... & -31.8 $\pm$\phantom{0}6.3 & 5.7 & ... & ... \\ 
01533081+5736078 & ... & 10.507 & 103 & 3/3 & 37 $\pm$\phantom{0}8\phantom{*} & \phantom{1}56 & 13 & ... & -52.9 $\pm$\phantom{0}6.4 & 6.7 & ... & ... \\ 
\noalign{\smallskip}\hline \noalign{\smallskip}
\end{tabular}
\begin{tablenotes}
\item[1] SIMBAD
\item[2] \citet{renson2009}
\end{tablenotes}
\end{threeparttable}
\end{table*}

\begin{table}
\centering
\caption{Atomic data for lines detected in the APOGEE spectra of HgMn stars. The {\mnii}, {\ci}, and hydrogen (Brackett series) data are from the Kurucz line list. The {\ceiii} data are from the Dream database.
}
\label{linelist}
\begin{tabular}{lccccr}
\noalign{\smallskip}\hline\hline
Ion & $\lambda_{\rm vac.}$ & log[$gf$] & $E_{\rm lo}$ & $E_{\rm hi}$ & $W_{\rm \lambda}/$ \\
    & ({\AA})              &           & (eV)         & (eV)         & $W_{\rm 15413}$     \\
\noalign{\smallskip}\hline\hline
{\mnii}  & 15368.065 & -1.002 & \phantom{1}9.863 & 10.670 & 0.19 \\
{\mnii}  & 15387.220 & -0.272 & \phantom{1}9.864 & 10.670 & 0.51 \\
{\mnii}  & 15412.667 & \phantom{-}0.237 & \phantom{1}9.865 & 10.670 & 1.00 \\
{\mnii}  & 15470.870 & -1.869 & \phantom{1}9.007 & \phantom{1}9.809 & 0.12 \\
{\mnii}  & 15586.570 & -0.558 & \phantom{1}9.862 & 10.658 & 0.30 \\
{\mnii}  & 15600.576 & -0.140 & \phantom{1}9.863 & 10.658 & 0.52 \\
{\mnii}  & 15620.314 & -0.031 & \phantom{1}9.864 & 10.658 & 0.70 \\
{\mnii}  & 15737.053 & -0.303 & \phantom{1}9.862 & 10.650 & 0.50 \\
{\mnii}  & 15746.494 & -0.257 & \phantom{1}9.862 & 10.650 & 0.50 \\
{\mnii}  & 15760.789 & -0.479 & \phantom{1}9.863 & 10.650 & 0.34 \\
{\ceiii} & 15961.157 & -1.120 & \phantom{1}0.000 & \phantom{1}0.777 & ... \\
{\ceiii} & 15964.928 & -1.660 & \phantom{1}0.815 & \phantom{1}1.591 & ... \\
{\ceiii} & 16133.170 & -0.920 & \phantom{1}0.388 & \phantom{1}1.156 & ... \\
{\ci}    & 16895.031 & \phantom{-}0.568 & \phantom{1}9.003 & \phantom{1}9.736 & ... \\
\noalign{\smallskip}\hline \noalign{\smallskip}
H-Br20   & 15196.005 & -1.489 & 12.749 & 13.564 & .... \\
H-Br19   & 15264.717 & -1.416 & 12.749 & 13.564 & .... \\
H-Br18   & 15345.991 & -1.339 & 12.749 & 13.564 & .... \\
H-Br17   & 15443.148 & -1.256 & 12.749 & 13.564 & .... \\
H-Br16   & 15560.708 & -1.167 & 12.749 & 13.564 & .... \\
H-Br15   & 15704.960 & -1.071 & 12.749 & 13.564 & .... \\
H-Br14   & 15884.888 & -0.966 & 12.749 & 13.564 & .... \\
H-Br13   & 16113.721 & -0.852 & 12.749 & 13.564 & .... \\
H-Br12   & 16411.681 & -0.725 & 12.749 & 13.564 & .... \\
H-Br11   & 16811.117 & -0.582 & 12.749 & 13.564 & .... \\
\noalign{\smallskip}\hline \noalign{\smallskip}
\end{tabular}
\end{table}
\normalsize

The search identified almost 500 stars with potential {\mnii} detections, and visual inspection of the resulting spectra confirmed $\sim400$ stars with {\mnii}. Five previously known HgMn stars (HD\,36662, 45975, 49606, 158704, and 182308) were included, and unlike the case of HD\,207857, the only metal lines in the associated spectra are {\mnii} (no {\ceiii}). Furthermore, 44 stars recovered by the search are classified as classical Ap/Bp stars in the literature, and for 40 of these, the {\mnii} lines are accompanied by {\ceiii} and either {\crii}, the $UL$, or both. We therefore defined the HgMn sample based on comparison of the equivalent width of the strong {\mnii}~15413\,{\AA} line versus the sum of equivalent widths of {\ceiii}\,15961 and 15965\,{\AA}, {\crii}~15370 and 15470\,{\AA}, and the $UL$ at 15231 and 16184\,{\AA}. This comparison is shown in the Figure~\ref{eqwplot}, with the dashed line at $y=60$\,m{\AA} indicating the fairly clean separation between HgMn stars and Ap/Bp stars. 

The three star symbols in the extreme lower left corner of Figure~\ref{eqwplot} correspond to stars with literature Mn-peculiar classifications that did not produce the $H$-band {\mnii} lines in the available spectra. For Renson 3040 (A0\,HgMn), the metal line content is limited to {\ci} and {\mgi}, which is typical of mid--early A-type stars. For HD\,51688 (B8\,SiMn), no metal lines are present in the spectra. For HD\,169027 (A0\,Mn), {\ceiii}~15961\,{\AA} and {\silii}~16911\,{\AA} are possibly present in lieu of the {\mnii} lines. 

The distinctions between HgMn stars versus apparently normal stars, Ap/Bp stars, and Am/Fm stars are further emphasized in Figure~\ref{specmontage}, which displays APOGEE spectra of two examples each of apparently normal stars, HgMn stars, Ap/Bp stars, and Am stars.  HgMn stars are most similar to the Ap/Bp stars in terms of line content, but only for the HgMn stars is the line content limited to {\mnii} with occasional {\ceiii} absorption and {\ci} emission. We are not aware of any examples of Ap/Bp stars with {\ci} in emission, but as will be discussed in Section~\ref{c1em}, it is a common trait of HgMn stars and exotic emission line stars. The spectrum of HD\,2559 in Figure~\ref{specmontage} demonstrates that the {\ci}~16895\,{\AA} line is also occasionally in emission for superficially normal late-B and early-A stars.

\subsection{Sample Parameters}
Figure~\ref{radecplot} presents some basic parameters of the sample, including the spatial, brightness, and temperature classification distributions. As expected, the vast majority of APOGEE HgMn stars are located near the Galactic midplane, having been observed in the grid of relatively-high-extinction Disk fields that have been the primary focus of the APOGEE survey. The sparsity of stars below declination $\simeq-30^{\circ}$ is caused by several factors, not only limited to the later start date of APOGEE-2 observations on the Du Pont Telescope. Another factor is the fact that during APOGEE-1 and part of APOGEE-2N, 35 telluric standard stars were observed per field. Partway into APOGEE-2 however, it was decided that the telluric correction would not suffer if that number was decreased to 15. Therefore, only 15 telluric standard stars per field have been observed during some of APOGEE-2N and all of APOGEE-2S. Further, the on-sky radii of the plug-plates for APOGEE-2S are typically 0.8$^{\circ}$, as compared to the 1.5$^{\circ}$ on-sky radii of APOGEE-2N plug-plates. The reduced field-of-view often leads to cooler stars being selected as telluric standard stars.

As demonstrated in the lower left panel of Figure~\ref{radecplot}, the APOGEE sample of HgMn stars is essentially an extension of the known sample to fainter magnitudes. The combined $V$ magnitude distributions of known versus new HgMn stars suggest that many more examples should exist between $7<V<8$. Although a large fraction of the APOGEE HgMn sample lacks any indication of spectral type in the literature, the lower right panel of Figure~\ref{radecplot} shows that the available spectral types are in line with expectations set by the previously known sample. With limited exceptions, the HgMn stars cover a remarkably narrow spectral type range of B7--A0. 

The APOGEE HgMn star sample is summarized in Table~\ref{startable}, which provides for each star the 2MASS designation, an alternative identifier, the 2MASS $H$ magnitude, the average S/N, the number of radial velocity (RV) measurements compared to the overall number of spectra, {\vsini}, the average equivalent widths ($W_{\lambda}$) of {\mnii}~15413\,{\AA}, {\ceiii}~15961\,{\AA}, and {\ci}~16895\,{\AA}, the average RV, the maximum difference between individual RVs (RV$_{\rm scat.}$), an indication of binarity, and if it exists, a literature spectral type from either SIMBAD or the chemically peculiar star catalog of \citet{renson2009}. 

In the Binary Type column of Table~\ref{startable}, ``VIS'' indicates that the star is known visual double or multiple system, ``SB1\,(lit.)'' indicates the star has been reported as a binary in the literature, and asterisks indicate that an orbital solution is reported in this paper (see Section~\ref{orbits}). Asterisks next to the {\vsini} values in Table~\ref{startable} indicate that {\vsini} was derived from optical spectra.

\begin{figure}
\centering 
\includegraphics[width=1.0\columnwidth]{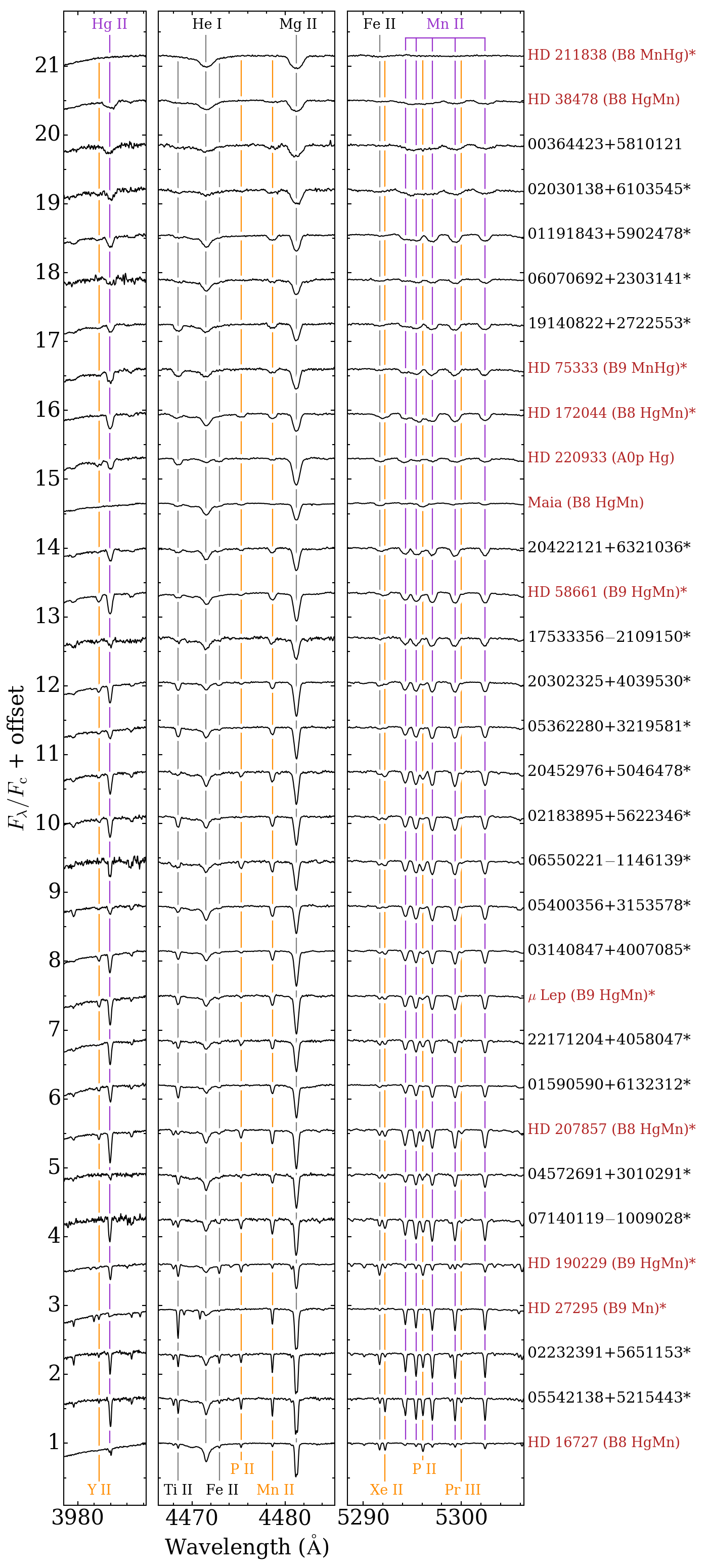}
\caption{Optical spectra of SB1 and apparently single HgMn stars, showing regions centered on {\hgii}~3984\,{\AA}, {\hei}~4471\,{\AA} and {\mgii}~4481\,{\AA}, and the {\mnii} lines between 5294--5302\,{\AA}. The spectra are sorted vertically by {\vsini}, going from 66\,{\kms} for HD\,211838 down to 2.8\,{\kms} for HD\,16727. Star names and 2MASS designations are given for the known and new HgMn stars, respectively, with literature spectral types given for the known HgMn stars and with asterisks indicating stars that are definite or suspected binaries. Grey vertical lines indicate features that appear in the spectra of normal B7--A0 stars, while purple and orange vertical lines indicate features that only appear in the spectra of chemically peculiar stars. } 
\label{specmontage2}
\end{figure}

\begin{figure}
\centering 
\includegraphics[width=1.0\columnwidth]{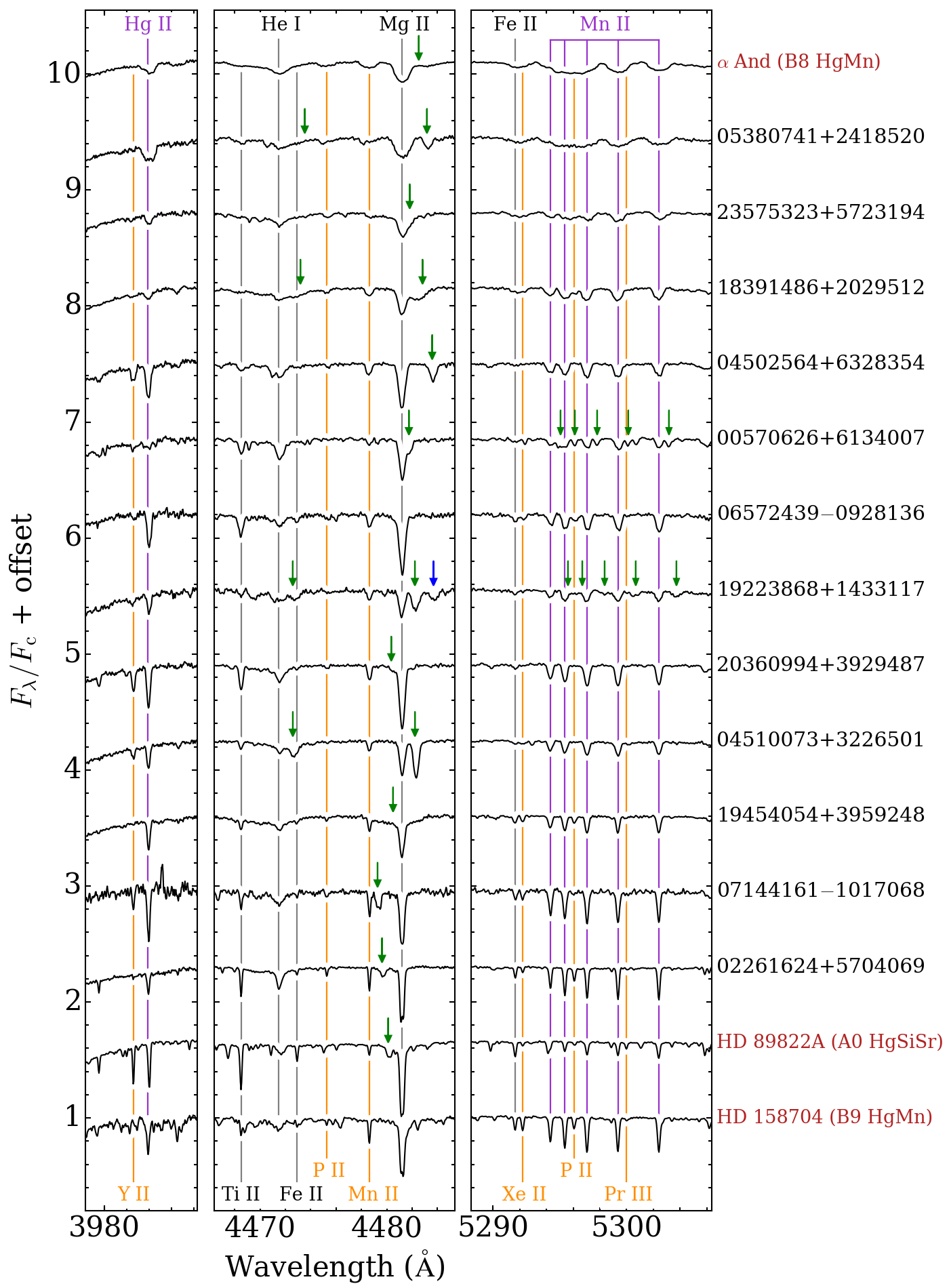}
\caption{The same as Figure~\ref{specmontage2} but for binary systems identified as multi-lined in either the optical or $H$-band. Small green arrows mark lines from the secondary stars in cases of $>20$\,{\kms} velocity separations of the components, and in the case of the SB3 2MASS\,J19223868+1433117, a blue arrow marks the {\mgii}~4481\,{\AA} line of the tertiary star.} 
\label{specmontage3}
\end{figure}

\subsection{$H$-band Linelist}
Atomic data for lines that can appear in the $H$-band spectra of HgMn stars are given in Table~\ref{linelist}. The Kurucz linelist\footnote{http://kurucz.harvard.edu/lineslists/} includes 145 {\mnii} lines, but only the strong lines (in terms of log$_{10}$ of the product of statistical weight, $g$, and oscillator strength, $f$) with lower energy levels around 10\,eV are detected. Although there are no Hg lines in the $H$-band, all of the new HgMn stars we observed in the optical exhibit {\hgii}, such that it is safe to say that a star whose $H$-band line content is limited to {\mnii} is almost certainly a bona fide HgMn star. The $W_{\lambda}/W_{\rm 15413}$ column of Table~\ref{linelist} gives the average equivalent width ratios of the {\mnii} lines over {\mnii}~15413\,{\AA}, based on measurements of the lines in spectra of stars with exceptionally strong {\mnii}. See \citet{choj2019} for a list of the empirical rest wavelengths of the $UL$.

\subsection{Optical Verification} \label{opticalspec}
Figures~\ref{specmontage2} and \ref{specmontage3} display selected regions of the optical follow-up spectra for new and previously known HgMn stars, allowing us to verify that the new finds are indeed HgMn stars as defined in the optical. Detection of {\hei} lines at strengths less than that of {\mgii}~4481\,{\AA} indicates that all of the stars fall in the B7--A0 temperature class range, as expected for HgMn stars. With the exceptions of HD\,211838 and Maia, the Hg peculiarity is confirmed for all of the stars via detection of {\hgii}~3984\,{\AA}, which is typically the strongest {\hgii} line in the optical spectra of HgMn stars \citep{cowley1975}. Depending on rotational velocity, dozens to hundreds of {\mnii} lines are readily identified in the spectra of the new HgMn stars, thus confirming the Mn peculiarity. The atlas of \citet{castelli2004b} was particularly useful for identifying weak heavy metal features, and indeed many of the stars exhibit anomalously strong lines from {\pii}, {\gaii}, {\yii}, {\zrii}, {\xeii}, {\ptii}, {\priii}, and {\ndiii}, all of which are known to be overabundant in HgMn stars \citep{ghazaryan2016}. The stars 
2MASS\,J02232391+5651153 and 2MASS\,J05542138+5215443 (near the bottom of Figure~\ref{specmontage2}) are perhaps the most extreme HgMn stars of our optical sample. We were able to identify $\sim$300 {\mnii} lines in the optical spectra of both stars, along with 83 {\pii} lines and 13 {\xeii} lines in the spectrum of 2MASS\,J05542138+5215443. 

\begin{table*}
\centering
\caption{Radial velocity data. Only the first 10 rows are shown (full version available online). }
\label{rvtable}
\begin{tabular}{cccccc}
\noalign{\smallskip}\hline\hline
2MASS       & HJD     & Inst. & RV$_{1}$ & RV$_{2}$ & RV$_{3}$  \\ 
Designation & 2.45E5+ &       & ({\kms}) & ({\kms}) & ({\kms})  \\
\noalign{\smallskip}\hline\hline
00091817+7022314 & 5850.780 & APOGEE & $-19.75\pm1.00$ & ... & ...  \\
00091817+7022314 & 5851.758 & APOGEE & $-19.77\pm1.00$ & ... & ...  \\
00091817+7022314 & 5869.680 & APOGEE & $-16.80\pm1.29$ & ... & ...  \\
00091817+7022314 & 5871.698 & APOGEE & $-17.28\pm1.94$ & ... & ...  \\
00091817+7022314 & 6173.835 & APOGEE & $-21.26\pm2.10$ & ... & ...  \\
00091817+7022314 & 6204.755 & APOGEE & $-20.20\pm2.75$ & ... & ...  \\
00091817+7022314 & 6224.689 & APOGEE & $-19.50\pm1.11$ & ... & ...  \\
00091817+7022314 & 6233.644 & APOGEE & $-16.61\pm4.32$ & ... & ...  \\
00091817+7022314 & 6234.656 & APOGEE & $-18.24\pm1.00$ & ... & ...  \\
00091817+7022314 & 6257.552 & APOGEE & $-19.05\pm1.57$ & ... & ...  \\
\noalign{\smallskip}\hline \noalign{\smallskip}
\end{tabular}
\end{table*}

The only stars represented in Figures~\ref{specmontage2} and \ref{specmontage3} for which no evidence of binarity or multiplicity has been either reported in the literature or in this paper are HD\,38478, 2MASS\,J00364423+5810121, HD\,220933, and Maia. All of the new HgMn stars were targeted for optical follow-up due to being likely or definite SB1s and SB2s. For most of the stars represented in Figure~\ref{specmontage3}, blue arrows mark the positions of the binary companions' {\mgii}~4481\,{\AA} lines and occasionally also of the companions' {\hei}~4471\,{\AA} lines. For 2MASS\,J06572439$-$0928136 and HD\,158704, the companions are separated from the primaries by $<20$\,{\kms}, such that the companion contributions to strong lines are thoroughly blended with the primary star lines. For $\alpha$\,And, the companion is difficult to detect owing to the rapid rotation of both stars, so the arrows simply mark the expected companion position based on the orbit presented by \citet{ryabchikova1999}. In the cases of the SB2 2MASS\,J00570626+6134007 and the SB3 2MASS\,J19223868+1433117, the primary and secondary stars both produce optical {\mnii} lines. The {\mgii}~4481\,{\AA} line of the tertiary in the latter system is marked with a blue arrow in Figure~\ref{specmontage3}. Most of these stars will be discussed in more detail in Section~\ref{binclass}.

\begin{figure}
\includegraphics[width=1.0\columnwidth]{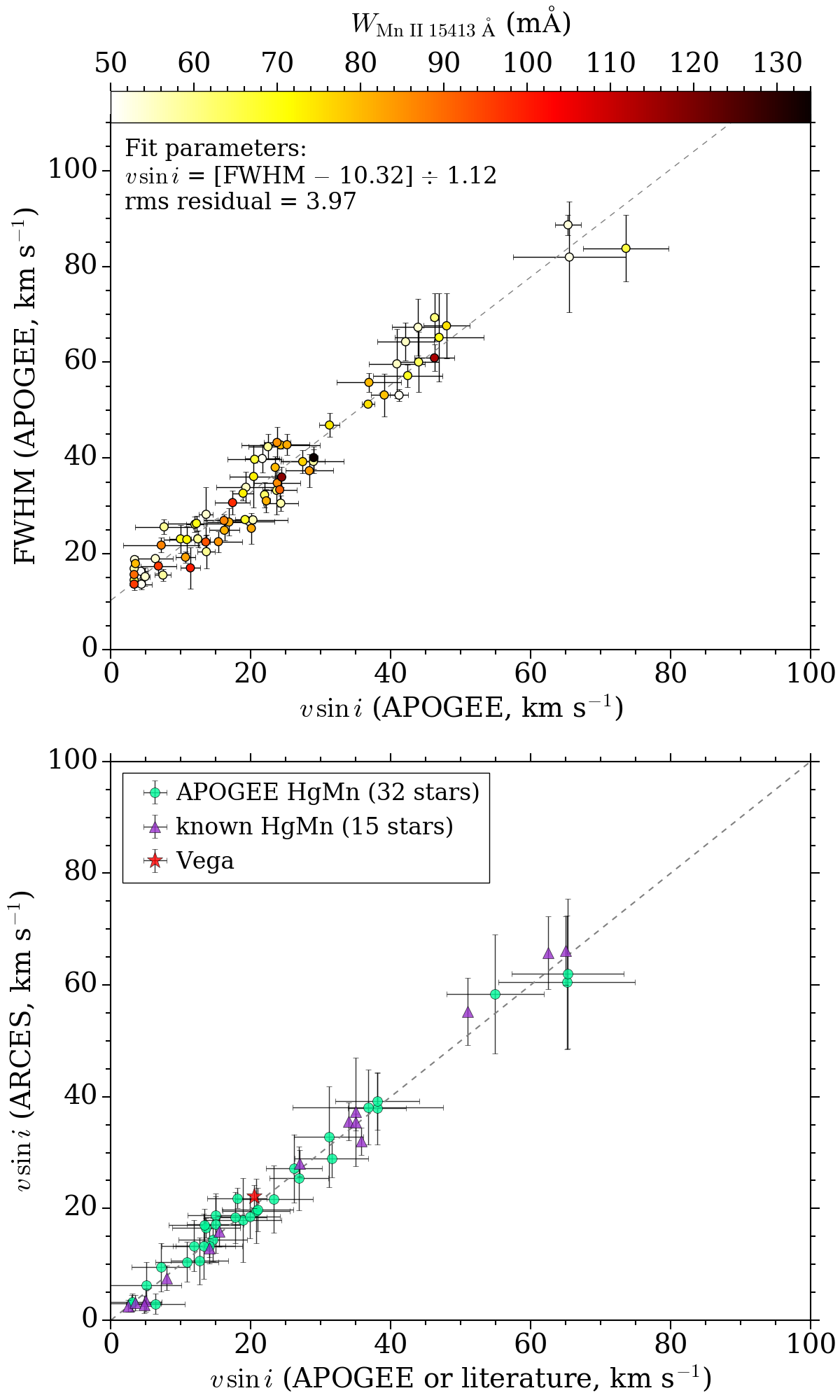}
\caption{(\emph{Top:}) A linear fit to the relation between goodness-of-fit {\vsini} results from the $iacob \,\,broad$ code versus Gaussian FWHMs of the same lines. Both quantities are the averages of multiple {\mnii} lines, and the points are colored by the equivalent width of {\mnii}~15413\,{\AA}. (\emph{Bottom:}) A comparison of {\vsini} measured from our optical spectra versus {\vsini} from the $H$-band data or the literature, with a dotted line at $y=x$. For the previously-known-but-not-observed-by-APOGEE-stars (purple triangles), the abscissa gives a literature {\vsini} estimate.}
\label{vsiniplot}
\end{figure}

\section{Radial Velocities} \label{rv}
Having established the HgMn sample, we proceeded to use the $H$-band {\mnii} lines to measure RVs by interactively fitting Gaussians to as many {\mnii} lines as possible from 1584/1709 APOGEE spectra (excluding some low S/N data) using the $splot$ program of IRAF. From this we obtained estimates of the line centers, FWHMs, and $W_{\lambda}$. The line position measurements from each spectrum were averaged and a barycentric correction was added to produce the heliocentric RVs quoted in this paper. For the errors, we simply adopted the standard deviation ($\sigma$) of the line-by-line measurements. In most cases, at least two {\mnii} lines (usually 15413\,{\AA} and 15620\,{\AA}) were resolved in each spectrum, but in spectra where only the strongest {\mnii} line (15413\,{\AA}) was resolved, we assign a constant RV error of 10\,{\kms}. For the optical follow-up sample, the same process described above was applied to numerous {\feii}, {\silii}, {\crii}, {\tiii}, {\mnii}, and {\oi} lines. The epoch-by-epoch RV measurements are provided in Table~\ref{rvtable}.

\section{Rotational Velocities} \label{vsini}
Rotational velocities ({\vsini}) were estimated based on the Gaussian FWHMs measured for all of the stars and a calibration derived from a sub-sample of 71 stars with particularly strong $H$-band {\mnii} lines ($W_{\rm Mn\,\,II\,\,15413\,\,\AA}>50$\,m{\AA}) and high-S/N spectra. The latter analysis was carried out using the IDL code $iacob \,\,broad$ \citep{diaz2014}, which estimates {\vsini} via two methods: identification of the first zero of the Fourier transform of an observed line profile, and reduction of $\chi^{2}$ via a goodness-of-fit (GOF) technique whereby a theoretical line profile convolved with rotational and macroturbulent line-broadening components is compared to the observed line profile. Due to considerably more scatter in a plot of the Gaussian FWHMs versus the Fourier transform {\vsini}, we adopted the GOF {\vsini}. The upper panel of Figure~\ref{vsiniplot} shows the associated linear fit to the FWHM and GOF {\vsini}. 

The relation between {\vsini} and FWHM derived from the sub-sample of 71 stars was then used to convert the Gaussian FWHM measurements to {\vsini} for the stars with weaker and/or noisier line profiles. For error estimates, the FWHM standard deviations were added in quadrature to the root mean square residual of 3.97,{\kms} of the liner fit in the upper panel of Figure~\ref{vsiniplot}. From this, we find an average of $<${\vsini}$>=26.8$\,{\kms} for the APOGEE sample, which is quite close to the average $<${\vsini}$>=30.6$\,{\kms} for the previously known HgMn stars. The ranges of {\vsini} are also similar, going from 3--88\,{\kms} for the APOGEE sample and 0--109\,{\kms} for the known sample.  

For the optical follow-up sample, we ran the $iacob \,\,broad$ code on minimally-blended lines that are relatively strong in the spectra of all of the stars. This included lines of {\feii} (4508, 4924, 5018, 5169\,{\AA}), {\mnii} (4479, 4756, 5559, 5571\,{\AA}), {\silii} (4128, 4131, 6347\,{\AA}), and {\oi} (7772, 7774, 7775\,{\AA}). The average GOF {\vsini} results are plotted in the lower panel of Figure~\ref{vsiniplot} against either the {\vsini} estimated from APOGEE spectra or else a literature value in the cases known HgMn stars which were not observed by APOGEE. The agreement is quite good, to within 5\,{\kms} in most cases. Due to the higher resolution and vastly greater number of available lines of the optical versus $H$-band spectra, Table~\ref{startable} reports the {\vsini} measured from optical spectra for the optical follow-up sample.

\section{Multiplicity} \label{binclass}
\begin{figure}
\includegraphics[width=1.0\columnwidth]{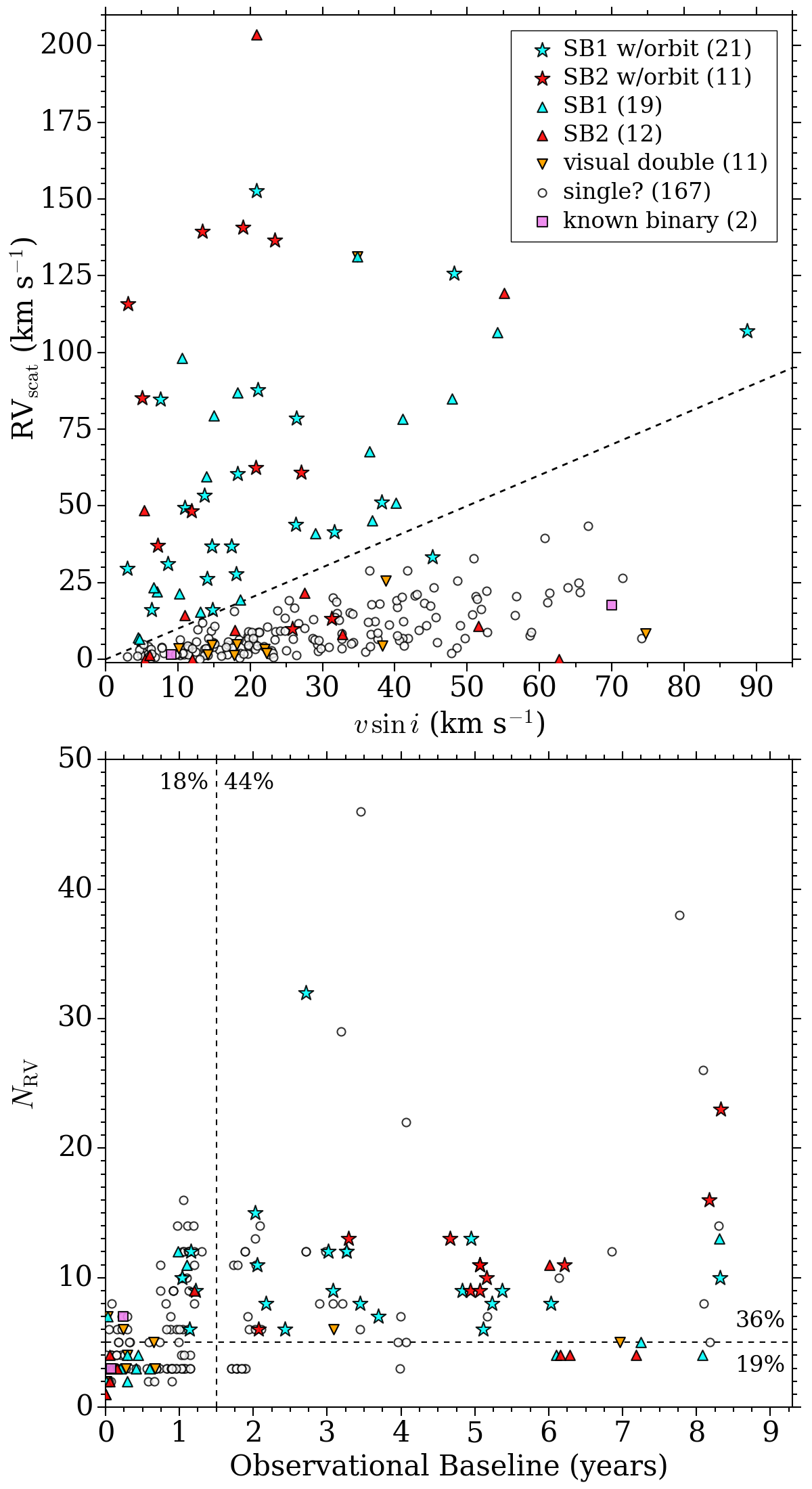}
\caption{(\emph{Top:}) Spectroscopic binary classification, whereby we consider everything with RV$_{\rm scat}>$\,{\vsini} a definite binary or multiple, with a few exceptions. The dashed line indicates RV$_{\rm scat}=$\,{\vsini}. (\emph{Bottom:}) Observational baseline versus number of RV measurements. The multiplicity fraction rises from 19\% to 45\% for stars with $>1.5$ years observational baselines, and it rises from 20\% to 37\% for stars with more than five spectra. }
\label{rvscatplot}
\end{figure}

HgMn stars are frequently or perhaps always member of binary or multiple star systems \citep[e.g.][]{dolk2003,scholler2010}, and the APOGEE sample is no different. This is demonstrated in the upper panel of Figure~\ref{rvscatplot}, which plots the maximum difference between individual RVs (RV$_{\rm scat}$) versus {\vsini}. Since it is possible for chemical spotting to make the line centers vary over the course of a rotational period, leading to measurable RV variability, we conservatively classify as definite binaries or multiples the stars for which RV$_{\rm scat}$ exceeds {\vsini} and/or the stars that are clear SB2s. We also make an exception for 2MASS\,J05400356+3153578 since a convincing SB1 orbital solution was found. Although we did not find RV variability that exceeds {\vsini} for the previously known HgMn stars HD\,45975 and HD\,207857, long-period binarity has been reported by \citet{andrews2017} and \citet{pourbaix2004}, respectively. These stars are represented by square symbols in Figure~\ref{rvscatplot}. The stars represented by downward-facing triangle symbols in Figure~\ref{rvscatplot} are known visual doubles that we did not classify as binaries based on the APOGEE spectra.

In addition to the above SB1 classification criteria, we visually examined all available spectra to check for evidence of lines with RV offsets compared to the {\mnii} lines. A total of 19 SB2 systems were found based on the APOGEE data alone, including two systems that exhibit spectral lines from at least three stars (SB3s). Another four systems were confirmed as SB2s based on optical spectra, despite the companion stars not being detected in the APOGEE spectra. 

Considering the 40 SB1 systems, 23 SB2s and SB3s, 11 visual double stars, and 2 previously known binaries, compared to the 165 stars whose RVs are not variable to within the projected {\vsini}, we arrive at very conservative multiplicity fraction estimate of $N_{\rm multiples}/N_{\rm total}=31$\% (76 total multiples). However, due to the numerous stars for which the APOGEE observations were too few and/or over too short timescales for detecting anything but very short-period binaries, 31\% is firmly a lower limit on the multiplicity fraction. As shown in the lower panel of Figure~\ref{rvscatplot}, the fraction increases toward longer observational baselines and/or more spectra, with for example the fraction being 18\% for baselines shorter than 1.5 years and 44\% for baselines longer than 1.5 years. Likewise, the multiplicity fraction is 19\% for stars with five or fewer spectra, but it rises to 36\% for stars with more than five spectra. Further observations of this sample will undoubtedly reveal numerous additional binaries/multiples, particularly those with long orbital periods. 

It is important to note that the sample also includes several stars that we classify as possibly single despite many APOGEE spectra taken over relatively long timescales. The star 2MASS\,J19451699+2237253 (HD\,344908, B8) is perhaps the best example, having been observed 38 times over a period of $\sim7.8$ years. Compared to the majority of the sample, this star's {\mnii} lines are quite strong and narrow, such that there is minimal uncertainty about the RVs. The maximum RV difference between epochs for this star is just 3.7\,{\kms}, implying that if a binary companion does exist, the orbital period is necessarily quite long or the companion mass quite low.

\begin{figure*}
\centering 
\includegraphics[width=1.0\textwidth]{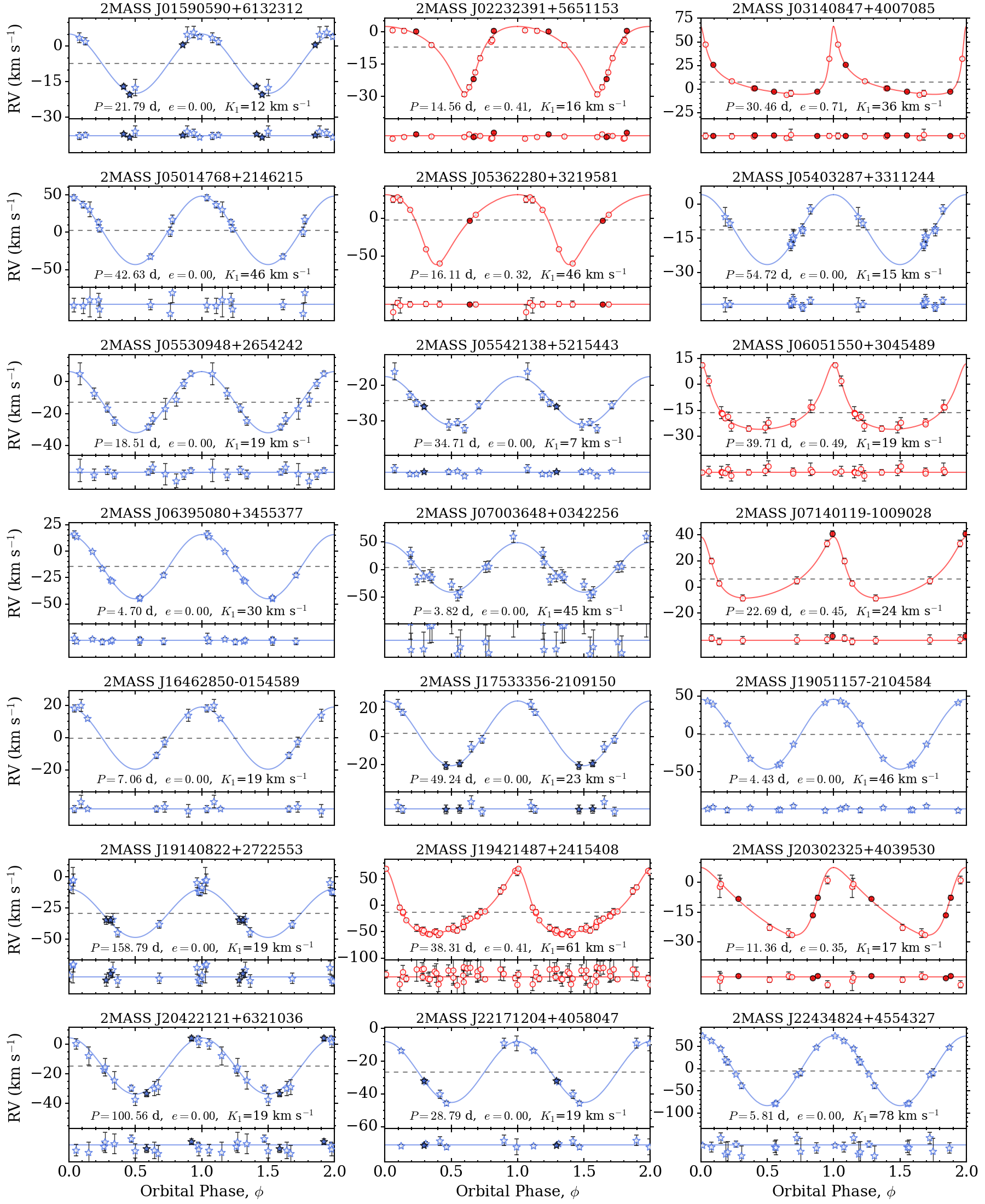}
\caption{Orbital solutions for SB1s. The orbits have been repeated for two phases to improve the sense of continuity. Star symbols (blue) indicate circular orbits, circle symbols (red) indicate eccentric orbits, and filled symbols pertain to RV measurements from optical spectra. Dashed horizontal lines in the larger panels indicate the systemic velocities ($\gamma$), and solid lines are fits to the data. The smaller panels show the residuals of the observations minus the fit, with the ordinates covering $\pm10$\,{\kms} and the horizontal lines indicating $y=0$.}
\label{sb1orbits}
\end{figure*} 

\begin{figure*}
\centering 
\includegraphics[width=1.0\textwidth]{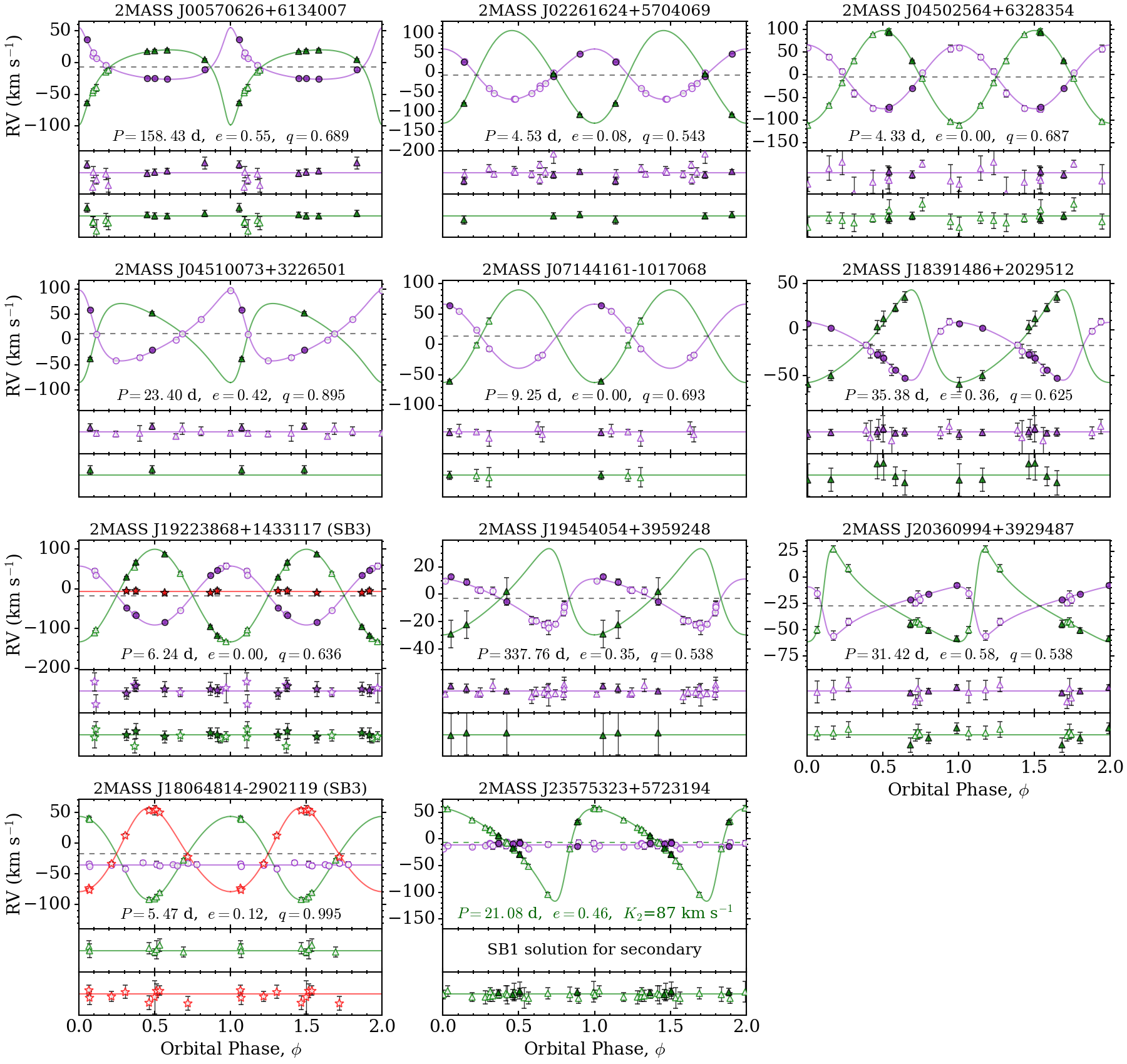}
\caption{Orbital solutions for SB2s and multiple systems. Circle symbols (purple) are the RVs of the HgMn stars, triangle symbols (green) are the RVs of the secondary stars, star symbols (orange) are the RVs of the tertiary in the two SB3s, and filled symbols pertain to RV measurements from optical spectra. In the larger panels, dashed horizontal lines indicate the systemic velocity ($\gamma$), and if applicable, solid horizontal lines indicate the average RV of the binary component not included in the orbital solution. The smaller panels are as described in Figure~\ref{sb1orbits}. For the SB3 2MASS\,J18064814$-$2902119, an SB2 orbital solution pertaining to the secondary (Ap) and tertiary (A/Am) components is shown, while for the SB3 2MASS\,J19223868+1433117, an SB2 orbital solution pertaining to the primary (HgMn) and tertiary (Am) components is shown. For 2MASS\,J23575323+5723194, an SB1 orbital solution pertaining to the Am component is shown.}
\label{sb2orbits}
\end{figure*}

\begin{landscape}
\begin{table}
\centering
\caption{Orbital solutions for SB1s.}
\label{sb1table}
\begin{tabular}{ccccccccccc}
\noalign{\smallskip}\hline\hline
2MASS ID & $N_{\rm RV}$ & $P$ & $T_{\rm max. RV}$ & $e$ & $\omega$ & $\gamma$ & $K_{\rm 1}$ & $a_{1}\, sin\,i$ & $f[M_{1},M_{2}$] & $\chi^{\rm 2}$ \\ 
         &               & (d) & (HJD-2.45E5)       &     & (deg.)   & ({\kms}) & ({\kms})       & (AU)     & ({\msun})        &                  \\ 
\noalign{\smallskip}\hline\hline
01590590+6132312 & \phantom{1}9 & $\phantom{3}21.794\pm0.007$ & $5791.904\pm0.864$ & $0.000\pm0.000$ & $\phantom{3}90.000\pm0.000$ & $\phantom{3}-7.416\pm0.402$ & $12.480\pm0.611$ & $0.025\pm0.001$ & $0.004\pm0.001$ & \phantom{3}6.081 \\ 
02232391+5651153 & 12 & $\phantom{3}14.565\pm0.001$ & $5805.879\pm0.235$ & $0.413\pm0.031$ & $201.647\pm3.774$ & $\phantom{3}-7.098\pm0.379$ & $15.514\pm0.517$ & $0.019\pm0.001$ & $0.004\pm0.001$ & 17.423 \\ 
03140847+4007085 & 10 & $\phantom{3}30.456\pm0.003$ & $5809.752\pm0.056$ & $0.720\pm0.048$ & $333.549\pm4.412$ & $\phantom{33}7.462\pm0.440$ & $36.576\pm5.721$ & $0.071\pm0.012$ & $0.052\pm0.029$ & \phantom{3}1.909 \\ 
05014768+2146215 & \phantom{1}8 & $\phantom{3}42.634\pm0.024$ & $5798.451\pm0.460$ & $0.000\pm0.000$ & $\phantom{3}90.000\pm0.000$ & $\phantom{33}2.570\pm1.647$ & $45.705\pm3.063$ & $0.179\pm0.012$ & $0.422\pm0.085$ & \phantom{3}3.118 \\ 
05362280+3219581 & \phantom{1}8 & $\phantom{3}16.109\pm0.001$ & $5796.707\pm0.113$ & $0.323\pm0.033$ & $148.052\pm3.275$ & $\phantom{3}-2.315\pm0.730$ & $46.344\pm1.040$ & $0.065\pm0.002$ & $0.141\pm0.018$ & \phantom{3}1.448 \\ 
05403287+3311244 & \phantom{1}9 & $\phantom{3}54.717\pm0.141$ & $5786.653\pm3.092$ & $0.000\pm0.000$ & $\phantom{3}90.000\pm0.000$ & $-11.192\pm1.024$ & $15.389\pm2.491$ & $0.077\pm0.013$ & $0.021\pm0.010$ & \phantom{3}3.665 \\ 
05530948+2654242 & 10 & $\phantom{3}18.507\pm0.024$ & $5796.129\pm1.485$ & $0.000\pm0.000$ & $\phantom{3}90.000\pm0.000$ & $-12.862\pm0.856$ & $19.105\pm1.210$ & $0.033\pm0.002$ & $0.013\pm0.003$ & \phantom{3}3.647 \\ 
05542138+5215443 & \phantom{1}8 & $\phantom{3}34.707\pm0.016$ & $5777.021\pm1.069$ & $0.000\pm0.000$ & $\phantom{3}90.000\pm0.000$ & $-24.348\pm0.568$ & $\phantom{3}6.800\pm1.065$ & $0.022\pm0.003$ & $0.001\pm0.001$ & \phantom{3}7.340 \\ 
06051550+3045489 & 15 & $\phantom{3}39.714\pm0.030$ & $5763.395\pm0.829$ & $0.488\pm0.034$ & $\phantom{33}8.508\pm6.811$ & $-16.222\pm0.506$ & $18.975\pm1.081$ & $0.060\pm0.004$ & $0.019\pm0.004$ & \phantom{3}3.559 \\ 
06395080+3455377 & \phantom{1}9 & $\phantom{33}4.701\pm0.000$ & $5798.665\pm0.037$ & $0.000\pm0.000$ & $\phantom{3}90.000\pm0.000$ & $-14.682\pm0.645$ & $30.296\pm0.762$ & $0.013\pm0.000$ & $0.014\pm0.001$ & \phantom{3}1.787 \\ 
07003648+0342256 & 12 & $\phantom{33}3.820\pm0.001$ & $5798.888\pm0.088$ & $0.000\pm0.000$ & $\phantom{3}90.000\pm0.000$ & $\phantom{33}3.773\pm3.177$ & $44.994\pm4.977$ & $0.016\pm0.002$ & $0.036\pm0.012$ & 13.570 \\ 
07140119$-$1009028 & \phantom{1}6 & $\phantom{3}22.690\pm0.009$ & $5784.093\pm1.042$ & $0.449\pm0.052$ & $\phantom{3}32.023\pm8.620$ & $\phantom{33}5.998\pm1.342$ & $23.531\pm1.350$ & $0.044\pm0.003$ & $0.022\pm0.006$ & \phantom{3}2.642 \\ 
16462850$-$0154589 & \phantom{1}6 & $\phantom{33}7.056\pm0.006$ & $5799.671\pm1.200$ & $0.000\pm0.000$ & $\phantom{3}90.000\pm0.000$ & $\phantom{3}-0.303\pm1.117$ & $19.308\pm2.442$ & $0.013\pm0.002$ & $0.005\pm0.002$ & \phantom{3}1.546 \\ 
17533356$-$2109150 & \phantom{1}6 & $\phantom{3}49.240\pm0.056$ & $5770.478\pm1.871$ & $0.000\pm0.000$ & $\phantom{3}90.000\pm0.000$ & $\phantom{33}2.490\pm1.749$ & $23.324\pm2.371$ & $0.106\pm0.011$ & $0.065\pm0.020$ & \phantom{3}1.876 \\ 
19051157$-$2104584 & \phantom{1}8 & $\phantom{33}4.433\pm0.000$ & $5800.286\pm0.082$ & $0.000\pm0.000$ & $\phantom{3}90.000\pm0.000$ & $\phantom{3}-0.495\pm0.377$ & $46.238\pm0.549$ & $0.019\pm0.000$ & $0.045\pm0.002$ & \phantom{3}5.314 \\ 
19140822+2722553 & 11 & $158.794\pm0.379$ & $5691.573\pm4.639$ & $0.000\pm0.000$ & $\phantom{3}90.000\pm0.000$ & $-29.031\pm1.510$ & $19.420\pm1.766$ & $0.283\pm0.026$ & $0.121\pm0.033$ & \phantom{3}8.389 \\ 
19421487+2415408 & 32 & $\phantom{3}38.311\pm0.013$ & $5776.054\pm0.298$ & $0.412\pm0.013$ & $\phantom{3}34.338\pm2.492$ & $-13.234\pm0.550$ & $60.766\pm0.922$ & $0.195\pm0.003$ & $0.673\pm0.043$ & 17.433 \\ 
20302325+4039530 & \phantom{1}9 & $\phantom{3}11.355\pm0.002$ & $5800.405\pm0.519$ & $0.346\pm0.053$ & $290.022\pm2.155$ & $-11.637\pm0.427$ & $17.125\pm0.420$ & $0.017\pm0.001$ & $0.005\pm0.001$ & \phantom{3}8.539 \\ 
20422121+6321036 & 13 & $100.563\pm0.171$ & $5748.877\pm3.348$ & $0.000\pm0.000$ & $\phantom{3}90.000\pm0.000$ & $-14.690\pm0.969$ & $19.122\pm0.966$ & $0.177\pm0.009$ & $0.073\pm0.011$ & \phantom{3}9.678 \\ 
22171204+4058047 & \phantom{1}7 & $\phantom{3}28.792\pm0.012$ & $5794.822\pm1.176$ & $0.000\pm0.000$ & $\phantom{3}90.000\pm0.000$ & $-26.608\pm1.170$ & $18.614\pm0.831$ & $0.049\pm0.002$ & $0.019\pm0.003$ & \phantom{3}3.180 \\ 
22434824+4554327 & 12 & $\phantom{33}5.808\pm0.000$ & $5799.736\pm0.056$ & $0.000\pm0.000$ & $\phantom{3}90.000\pm0.000$ & $\phantom{3}-4.978\pm1.057$ & $78.317\pm1.228$ & $0.042\pm0.001$ & $0.289\pm0.014$ & \phantom{3}6.719 \\ 
\noalign{\smallskip}\hline \noalign{\smallskip}
\end{tabular}
\end{table}
\end{landscape}

\begin{landscape}
\begin{table}
\centering
\caption{Orbital solutions for SB2s and SB3s.}
\label{sb2table}
\begin{tabular}{ccccccccccc}
\noalign{\smallskip}\hline\hline
2MASS ID & $N_{\rm RV}$ & $P$ & $T_{\rm max. RV}$ & $e$ & $\omega$ & $\gamma$ & $K_{\rm 1,2}$ & $a_{1,2}\, sin\,i$ & $M_{1,2}\,sin\,i^{3}$ & $\chi^{\rm 2}$ \\ 
         &               & (d) & (HJD-2.45E5)       &     & (deg.)   & ({\kms}) & ({\kms})       & (AU)               & ({\msun})             &                 \\ 
\noalign{\smallskip}\hline\hline
00570626+6134007A & 10 & $158.426\pm0.085$ & $5853.884\pm1.769$ & $0.551\pm0.551$ & $346.451\pm2.515$ & $\phantom{2}-7.448\pm0.430$ & $\phantom{1}40.703\pm2.886$ & $0.495\pm0.037$ & $5.621\pm0.978$ & 29.580 \\ 
00570626+6134007B & 10 & $158.426\pm0.085$ & $5853.884\pm1.769$ & $0.551\pm0.551$ & $346.451\pm2.515$ & $\phantom{2}-7.448\pm0.430$ & $\phantom{2}59.114\pm3.998$ & $0.718\pm0.052$ & $3.870\pm0.649$ & 29.580 \\ 
\hline
02261624+5704069A & 13 & \phantom{33}$4.527\pm0.000$ & $5798.103\pm0.071$ & $0.080\pm0.080$ & $61.463\pm4.823$ & $\phantom{2}-6.705\pm0.428$ & $\phantom{1}64.129\pm0.436$ & $0.027\pm0.000$ & $1.823\pm0.056$ & 19.628 \\ 
02261624+5704069B & \phantom{1}3 & \phantom{33}$4.527\pm0.000$ & $5798.103\pm0.071$ & $0.080\pm0.080$ & $61.463\pm4.823$ & $\phantom{2}-6.705\pm0.428$ & $118.139\pm1.548$ & $0.049\pm0.001$ & $0.990\pm0.020$ & 19.628 \\  
\hline
04502564+6328354A & 11 & \phantom{33}$4.334\pm0.000$ & $5798.698\pm0.019$ & $0.000\pm0.000$ & $90.000\pm0.000$ & $\phantom{2}-4.634\pm0.705$ & $\phantom{1}70.442\pm1.548$ & $0.028\pm0.001$ & $1.378\pm0.043$ & 21.919 \\ 
04502564+6328354B & 11 & \phantom{33}$4.334\pm0.000$ & $5798.698\pm0.019$ & $0.000\pm0.000$ & $90.000\pm0.000$ & $\phantom{2}-4.634\pm0.705$ & $102.545\pm1.214$ & $0.041\pm0.000$ & $0.946\pm0.040$ & 21.919 \\ 
\hline
04510073+3226501A & \phantom{1}9 & \phantom{3}$23.405\pm0.002$ & $5779.814\pm0.217$ & $0.417\pm0.417$ & $55.914\pm1.786$ & $\phantom{2}11.356\pm0.463$ & $\phantom{1}70.006\pm0.765$ & $0.137\pm0.002$ & $3.126\pm0.288$ & 12.952 \\ 
04510073+3226501B & \phantom{1}2 & \phantom{3}$23.405\pm0.002$ & $5779.814\pm0.217$ & $0.417\pm0.417$ & $55.914\pm1.786$ & $\phantom{2}11.356\pm0.463$ & $\phantom{2}78.182\pm3.372$ & $0.153\pm0.007$ & $2.799\pm0.154$ & 12.952 \\ 
\hline
07144161$-$1017068A & \phantom{1}6 & \phantom{33}$9.251\pm0.004$ & $5800.291\pm0.845$ & $0.000\pm0.000$ & $90.000\pm0.000$ & $\phantom{2}13.587\pm0.912$ & $\phantom{1}52.587\pm2.503$ & $0.045\pm0.002$ & $1.202\pm0.175$ & \phantom{3}1.275 \\ 
07144161$-$1017068B & \phantom{1}3 & \phantom{33}$9.251\pm0.004$ & $5800.291\pm0.845$ & $0.000\pm0.000$ & $90.000\pm0.000$ & $\phantom{2}13.587\pm0.912$ & $\phantom{2}75.916\pm4.876$ & $0.065\pm0.004$ & $0.832\pm0.096$ & \phantom{3}1.275 \\ 
\hline
18391486+2029512A & 13 & \phantom{3}$35.378\pm0.013$ & $5821.247\pm1.349$ & $0.357\pm0.357$ & $235.924\pm7.275$ & $-17.055\pm0.966$ & $\phantom{1}31.385\pm1.804$ & $0.095\pm0.006$ & $1.000\pm0.213$ & \phantom{3}4.338 \\ 
18391486+2029512B & \phantom{1}6 & \phantom{3}$35.378\pm0.013$ & $5821.247\pm1.349$ & $0.357\pm0.357$ & $235.924\pm7.275$ & $-17.055\pm0.966$ & $\phantom{2}50.249\pm4.562$ & $0.153\pm0.014$ & $0.625\pm0.099$ & \phantom{3}4.338 \\ 
\hline
19223868+1433117A & 11 & \phantom{33}$6.235\pm0.000$ & $5797.070\pm0.026$ & $0.000\pm0.000$ & $90.000\pm0.000$ & $-17.065\pm0.593$ & $\phantom{1}74.187\pm1.026$ & $0.043\pm0.001$ & $2.744\pm0.067$ & 10.823 \\ 
19223868+1433117B & \phantom{1}5 & ... & ... & ... & ... & ...& ... & ... & ... & ... \\ 
19223868+1433117C & 11 & \phantom{33}$6.235\pm0.000$ & $5797.070\pm0.026$ & $0.000\pm0.000$ & $90.000\pm0.000$ & $-17.065\pm0.593$ & $116.634\pm1.149$ & $0.067\pm0.001$ & $1.745\pm0.048$ & 10.823 \\ 
\hline
19454054+3959248A & 16 & $337.756\pm1.956$ & $5810.319\pm4.003$ & $0.352\pm0.352$ & $243.553\pm17.223$ & $\phantom{2}-3.093\pm0.703$ & $\phantom{1}16.937\pm1.706$ & $0.492\pm0.071$ & $2.115\pm0.823$ & 12.284 \\ 
19454054+3959248B & \phantom{1}3 & $337.756\pm1.956$ & $5810.319\pm4.003$ & $0.352\pm0.352$ & $243.553\pm17.223$ & $\phantom{2}-3.093\pm0.703$ & $\phantom{2}31.467\pm3.093$ & $0.914\pm0.130$ & $1.139\pm0.428$ & 12.284 \\ 
\hline
20360994+3929487A & \phantom{1}9 & \phantom{3}$31.422\pm0.012$ & $5774.023\pm0.576$ & $0.582\pm0.582$ & $112.929\pm3.548$ & $-27.390\pm0.666$ & $\phantom{1}23.770\pm1.990$ & $0.056\pm0.005$ & $0.358\pm0.057$ & 11.305 \\ 
20360994+3929487B & \phantom{1}9 & \phantom{3}$31.422\pm0.012$ & $5774.023\pm0.576$ & $0.582\pm0.582$ & $112.929\pm3.548$ & $-27.390\pm0.666$ & $\phantom{2}44.203\pm2.145$ & $0.104\pm0.006$ & $0.192\pm0.036$ & 11.305 \\ 
\hline
18064814$-$2902119A &  11       & ...    & ...     & ...     & ...     & ...    & ...    & ...    & ...      \\ 
18064814$-$2902119B & \phantom{1}7 & \phantom{33}$5.469\pm0.000$ & $5789.682\pm0.210$ & $0.116\pm0.116$ & $151.162\pm8.748$ & $-17.647\pm0.793$ & $\phantom{2}68.402\pm1.515$ & $0.034\pm0.001$ & $0.707\pm0.034$ & \phantom{3}7.216 \\ 
18064814$-$2902119C & \phantom{1}9 & \phantom{33}$5.469\pm0.000$ & $5789.682\pm0.210$ & $0.116\pm0.116$ & $151.162\pm8.748$ & $-17.647\pm0.793$ & $\phantom{2}68.402\pm1.515$ & $0.034\pm0.001$ & $0.704\pm0.031$ & \phantom{3}7.216 \\ 
\hline
23575323+5723194A & 23 &  ...       & ...    & ...     & ...     & ...     & ...    & ...    & ...    & ...      \\ 
23575323+5723194B & 23 & \phantom{3}$21.080\pm0.001$ & $5806.240\pm0.084$ & $0.456\pm0.456$ & $234.674\pm1.625$ & $\phantom{2}-7.197\pm0.930$ & $\phantom{2}86.620\pm2.911$ & $0.149\pm0.005$ & ...    & \phantom{3}4.260 \\ 
\noalign{\smallskip}\hline \noalign{\smallskip}
\end{tabular}
\end{table}
\end{landscape}

\begin{table*}
\centering
\scriptsize
\caption{Summary of double-lined and triple-lined spectroscopic binaries. Mass ratios ($q=M_{2}/M_{1}$) are given for the systems with orbital solutions, and flux ratios ($F_{2}/F_{1}$) of {\mgii}~4481\,{\AA} are given for the systems with optical spectra. The {\vsini} [1], {\vsini} [2], and {\vsini} [3] columns give rotational velocity estimates for the primary, secondary, and tertiary stars where applicable and where possible to make such estimates.
}
\label{sb2infotable}
\begin{tabular}{clccccccl}
\noalign{\smallskip}\hline\hline
            &           & Orbital & Mass            & {\mgii}         &               &               &               &       \\
2MASS       & Companion & Period  & Ratio           & Ratio           & {\vsini} [1]  & {\vsini} [2]  & {\vsini} [3]  & Note  \\
Designation & Type(s)   & (days)  & ($M_{2}/M_{1}$) & ($F_{2}/F_{1}$) & ({\kms})      & ({\kms})      & ({\kms})      &       \\
\noalign{\smallskip}\hline\hline
09432220$-$5348264 & HgMn                & ...             & ...              & ...                 & $\phantom{0}5\pm\phantom{0}5$ & $\phantom{0}5\pm\phantom{0}5$   & ...                    & ... \\
19223868+1433117   & BpMn, Am            & \phantom{15}6.2 & 0.636            & 0.967, 0.438        & $18\pm\phantom{0}2$           & $17\pm\phantom{0}2$             & $22\pm\phantom{0}5$    & SB3; SB2 orbit for HgMn+Am components \\
00570626+6134007   & Ap/Bp               & 158.4           & 0.689            & 0.245               & $20\pm\phantom{0}1$           & $\phantom{0}8\pm\phantom{0}2$   & ...                    & ... \\
02060290+5009136   & Ap/Bp               & ...             & ...              & ...                 & $34\pm\phantom{0}4$           & $29\pm10$                       & ...                    & ... \\
18064814$-$2902119 & Ap/Bp, A/Am         & ...             & ...              & ...                 & $26\pm\phantom{0}5$           & $14\pm\phantom{0}6$             & $15\pm\phantom{0}5$    & SB3; SB2 orbit for Ap+A/Am components \\
02261624+5704069   & Am                  & \phantom{15}4.5 & 0.543            & 0.203               & $\phantom{0}3\pm\phantom{0}1$ & $\phantom{0}4\pm\phantom{0}1$   & ...                    & companion not detected in $H$-band \\
04502564+6328354   & Am                  & \phantom{15}4.3 & 0.687            & 0.352               & $22\pm\phantom{0}2$           & $16\pm\phantom{0}3$             & ...                    & ... \\
05380741+2418520   & Am                  & ...             & ...              & 0.271               & $58\pm\phantom{0}3$           & $15\pm\phantom{0}4$             & ...                    & ... \\
07144161$-$1017068 & Am                  & \phantom{15}9.3 & 0.694            & 0.328               & $\phantom{0}6\pm\phantom{0}3$ & $\phantom{0}6\pm\phantom{0}3$   & ...                    & ... \\
17314436$-$2616112 & Am                  & ...             & ...              & 0.134               & $\phantom{0}3\pm\phantom{0}1$ & $13\pm\phantom{0}2$             & ...                    & ... \\
20360994+3929487   & Am                  & \phantom{1}31.4 & 0.536            & 0.115               & $13\pm\phantom{0}2$           & $11\pm\phantom{0}3$             & ...                    & ... \\
23575323+5723194   & Am                  & ...             & ...              & 0.227               & $33\pm\phantom{0}2$           & $11\pm\phantom{0}1$             & ...                    & ... \\
04510073+3226501   & late-B              & \phantom{1}23.4 & 0.895            & 1.100               & $13\pm\phantom{0}2$           & $10\pm\phantom{0}3$             & ...                    & companion not detected in $H$-band \\
18391486+2029512   & late-B              & \phantom{1}35.4 & 0.625            & 0.787               & $25\pm\phantom{0}7$           & $70\pm10$                       & ...                    & companion not detected in $H$-band \\
19454054+3959248   & late-B              & 337.8           & 0.538            & $\sim$1             & $\phantom{0}9\pm\phantom{0}1$ & $\sim200$                       & ...                    & companion not detected in $H$-band \\
02203397+6023494   & A/Am                & ...             & ...              & ...                 & $27\pm10$                     & $16\pm\phantom{0}7$             & ...                    & multiple system? \\
04315476+2138086   & A/Am                & ...             & ...              & ...                 & $62\pm\phantom{0}8$           & $62\pm15$                       & ...                    & ... \\
06402199$-$0322092 & A/Am                & ...             & ...              & ...                 & $51\pm\phantom{0}9$           & $23\pm\phantom{0}7$             & ...                    & multiple system? Long-period SB2? \\
06561603$-$2839053 & A/Am                & ...             & ...              & ...                 & $\phantom{0}9\pm\phantom{0}4$ & $22\pm\phantom{0}5$             & ...                    & ... \\
06572439$-$0928136 & A/Am                & ...             & ...              & ...                 & $28\pm\phantom{0}4$           & $15\pm\phantom{0}6$             & ...                    & ... \\
14482274$-$5959316 & A/Am                & ...             & ...              & ...                 & $21\pm11$                     & $18\pm\phantom{0}9$             & ...                    & ... \\
16133258$-$6019519 & A/Am                & ...             & ...              & ...                 & $11\pm\phantom{0}4$           & ...                             & ...                    & tentative detection in all APOGEE spectra \\
19294384$-$1801510 & A/Am                & ...             & ...              & ...                 & $12\pm\phantom{0}5$           & $18\pm\phantom{0}7$             & ...                    & ... \\
\noalign{\smallskip}\hline \noalign{\smallskip}
\end{tabular}
\end{table*}
\normalsize

\subsection{Binary Orbital Solutions} \label{orbits}
In the cases of RV-variable stars with at least six RV measurements, we attempted to determine Keplerian orbital parameters using the IDL code $rvfit$ \citep{marzoa2015}. This code utilizes an Adaptive Simulated Annealing algorithm to determine orbital parameters for both SB1s and SB2s given inputs of heliocentric Julian dates (HJD), RVs, and RV errors. For SB1s, the code outputs seven parameters including the orbital period ($P$), time of periastron passage ($T_{P}$), orbital eccentricity ($e$), the argument of the periastron ($\omega$), the systemic radial velocity ($\gamma$), the radial velocity semi-amplitude of the primary star ($K_{1}$). These parameters can then be used to determine the mass function ($f(m)$) and semi-major axis of the primary ($a_{1}\,sin\,i$). For SB2s, the radial velocity semi-amplitude of the secondary star ($K_{2}$) is another output, allowing for calculation of semi-major axes ($a_{1,2}\,sin\,i$, $a\,sin\,i$), dynamical masses ($M_{1,2}\sin\,i\,^{3}$), and the mass ratio ($q=M_{2}/M{1}$). Any of the output parameters from $rvfit$ can be fixed to a particular value or constrained by upper and lower limits in an input configuration file. In all cases we began with very loose parameter constraints and with $e$ as a free parameter, but in cases where the value of $e$ became comparable to its error, we forced circular orbits by setting $e=0$. Since $T_{P}$ is undefined for circular orbits, we instead provide an epoch of maximum RV ($T_{\rm max.\,RV}$) regardless of $e$. 

Orbital solutions for 21 SB1 systems and for at least one component in 11 SB2 and SB3 systems are presented in Figures~\ref{sb1orbits} and \ref{sb2orbits}, with open symbols pertaining to RVs measured from APOGEE spectra and filled symbols pertaining to RVs measured from optical spectra. The periods range from 4--338 days, with only six systems having periods longer than 50 days. Non-zero eccentricities are found in 15 cases, led by the $e=0.72$ of the SB1  2MASS\,J03140847+4007085. 

The minimum number of RV measurements for which we were able to find a reasonable orbital solution was six in the cases of the SB1s 2MASS\,J07140119$-$1009028, 2MASS\,J16462850-0154589, and 2MASS\,J17533356-2109150, and for the SB2 2MASS\,J07144161-1017068. Only for 2MASS\,J07144161-1017068 is the solution stable however, thanks to a group of five APOGEEspectra separated by just 63 days. For 2MASS\,J16462850-0154589, numerous periods between 6--60 days are just as reasonable as the result shown in Figure~\ref{sb1orbits}, and the orbit may also be slightly eccentric. For 2MASS\,J17533356-2109150, another common solution beyond the one shown in Figure~\ref{sb1orbits} has $P=171$ days. The orbits of these stars will likely be revised by additional observations.

The full orbital parameters of the SB systems presented in Figures~\ref{sb1orbits} and \ref{sb2orbits} are given in Tables~\ref{sb1table} and \ref{sb2table}.

\subsection{Binary Companions}
The majority of the $H$-band-detected companions were evident in the $NSL$ (namely, {\ci}, {\mgi}, {\sili}, and occasionally {\si}), but not in the {\ceiii} lines, indicating that they are either A or Am stars (see the Am star spectra in Figure~\ref{specmontage}). As these stars are significantly fainter than the HgMn primaries, the features are typically quite weak, often only becoming evident after smoothing the spectra. For Am stars, the {\ceiii} lines are always significantly weaker than {\mgi} and {\sili} such that we do not expect to be able to distinguish a normal A-type companion from an Am companion in the APOGEE spectra of HgMn SB2s. The {\ceiii} lines would likely be below our detection threshold. We therefore refer to the companions in these systems simply as ``A/Am'' due to the ambiguity, but in reality it would not be surprising if the majority of the companions were indeed chemically peculiar stars. 

On the other hand, at least 11 of the SB2 systems definitely involve chemically peculiar companions. This includes four cases of {\ceiii} lines that are offset from {\mnii}, if not clearly varying in anti-phase with respect to {\mnii}, and one case of multiple sets of {\mnii} lines in the APOGEE spectra. In another six cases, we classify the companions as Am stars based on detection of corresponding {\srii} and {\baii} lines in optical follow-up spectra \citep[known to be overabundant for Am stars; see][]{ghazaryan2018}. 

Table~\ref{sb2infotable} summarizes the multi-lined systems, providing the companion types and {\vsini} along with the mass ratios for systems with orbital solutions, and {\mgii}~4481\,{\AA} flux ratios for systems with optical spectra. For systems with optical spectra in which the companion lines were sufficiently separated from the primary lines, the companion {\vsini} estimates were determined using the $iacob \,\,broad$ code. Otherwise, the calibration presented in Figure~\ref{rvscatplot} was used. In many cases, the binary components have nearly or definitely identical {\vsini}, and in only three cases do we find one or more components with {\vsini}\,$>$30\,{\kms}. 

As noted in Section~\ref{binclass}, the companion stars of four SB2s were detected in the optical but not in the $H$-band. Two of these companions are late-B, rapid rotators, such that we would not expect to see corresponding metal lines in the $H$-band (see the upper two spectra in Figure~\ref{specmontage}.) Another two of the optical SB2s have cool, faint companions whose $H$-band lines are apparently too weak for detection in APOGEE spectra. Hence, although no convincing evidence is found for lines from companion star(s) in the APOGEE spectra of the SB1s represented in Figure~\ref{sb1orbits}, optical spectroscopy will undoubtedly reveal additional SB2s among them, with companions that are too faint, hot, or rapidly rotating to be detected in the $H$-band.

In the following, we discuss the SB2 and SB3 systems individually.

\subsection{Binaries with HgMn or Ap/Bp Companions}

\noindent\textbf{2MASS\,J00570626+6134007 = HD\,5429} is the clearest available example of an HgMn+Ap/Bp binary, making it the second such system known in addition to HD\,161701 \citep{gonzalez2014}. The {\ceiii} lines from the secondary star are quite obvious in the five APOGEE spectra as are numerous lines in our five optical spectra of the system, allowing us to derive the orbital solution shown in Figure~\ref{sb2orbits}. With $P=159$ days and $e=0.55$, this is one of the widest and most eccentric binaries of the APOGEE HgMn sample. The weakness of the {\hei} lines from the companion in the optical spectra indicates that it is either an late-B or early-A type star, and this is also suggested by the mass ratio ($M_{2}/M_{1}=0.689$) obtained from the orbital solution. The orbit also suggests a minimum mass of 5.6\,$M_{\odot}$ for the primary, which places it among the most massive HgMn stars. 

\begin{figure}
\centering 
\includegraphics[width=1.0\columnwidth]{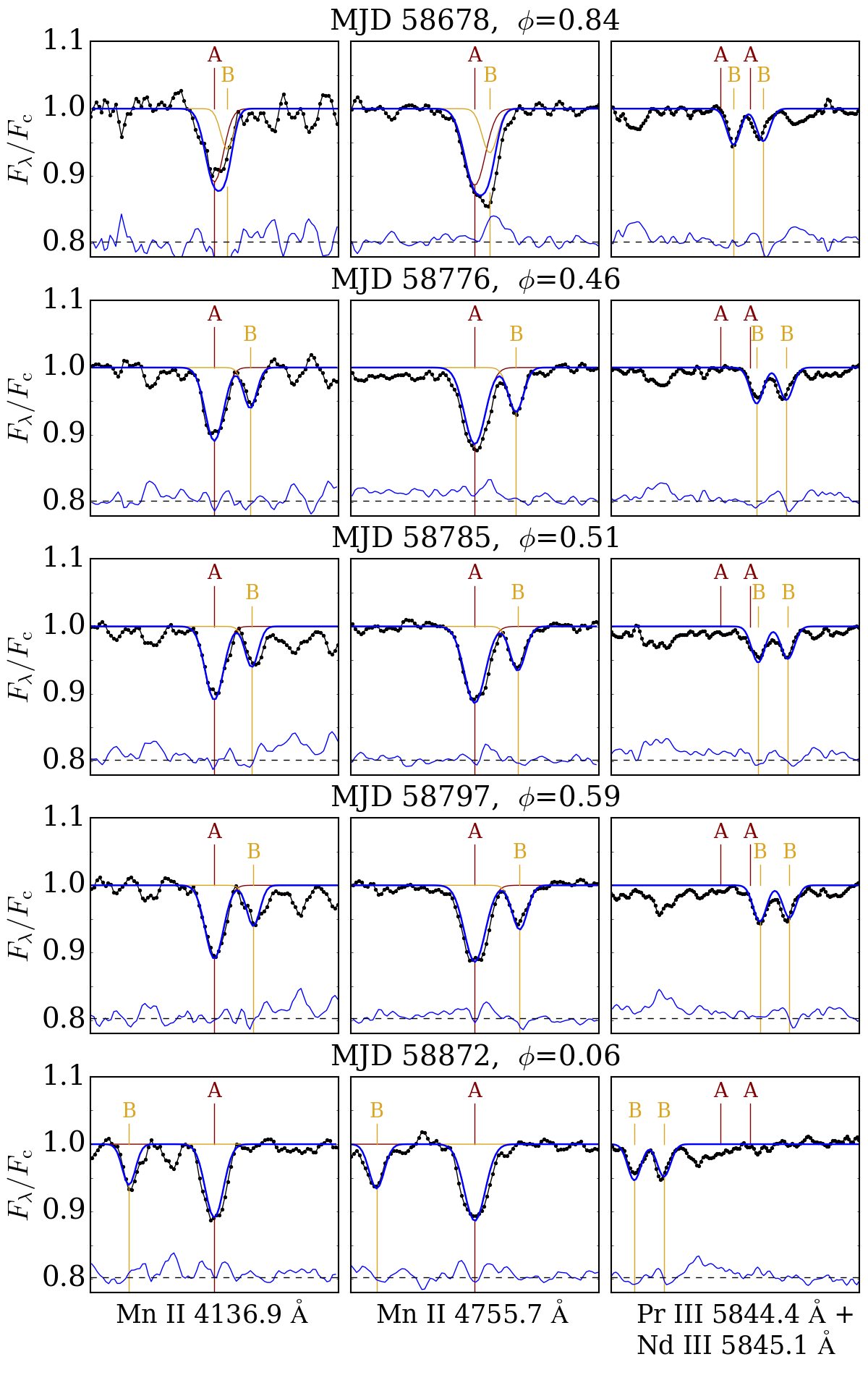}
\caption{Optical spectra of the HgMn+Ap/Bp binary 2MASS\,J00570626+6134007 (HD\,5429), highlighting the {\mnii} lines that are produced by both stars and the rare earth lines (e.g. {\priii} and {\ndiii}) that are produced by the Ap star (B) but not by the HgMn star (A). The spectra are shown in the rest frame of the HgMn star, and with individual Gaussians fits to the lines of each star (red and orange) as well as the sum of contributions from both stars (blue). Residuals are shown along the bottom of each panel, shifted upward artificially with the dashed lines indicating $y=0$. The quoted orbital phases pertain to the orbital solution shown in Figure~\ref{sb2orbits}.}
\label{fig00570626}
\end{figure}

Figure~\ref{fig00570626} shows selected regions of the optical spectra, with the left columns demonstrating that both stars produce the low-energy {\mnii} lines that are numerous in the optical, and with the right column demonstrating that only the Ap/Bp star contributes to doubly-ionized rare earth lines like {\priii} and {\ndiii}. The {\hgii}~3984\,{\AA} line of the HgMn primary is among the weakest of our optical sample (see Figure~\ref{specmontage3}), but it is certainly present and without a counterpart from the Ap/Bp star. Instead, the {\feii} and {\crii} lines of the Ap/Bp star are overly strong, further suggesting that it is a magnetic chemically peculiar star.

\vspace{0.1in}

\begin{figure}
\centering 
\includegraphics[width=1.0\columnwidth]{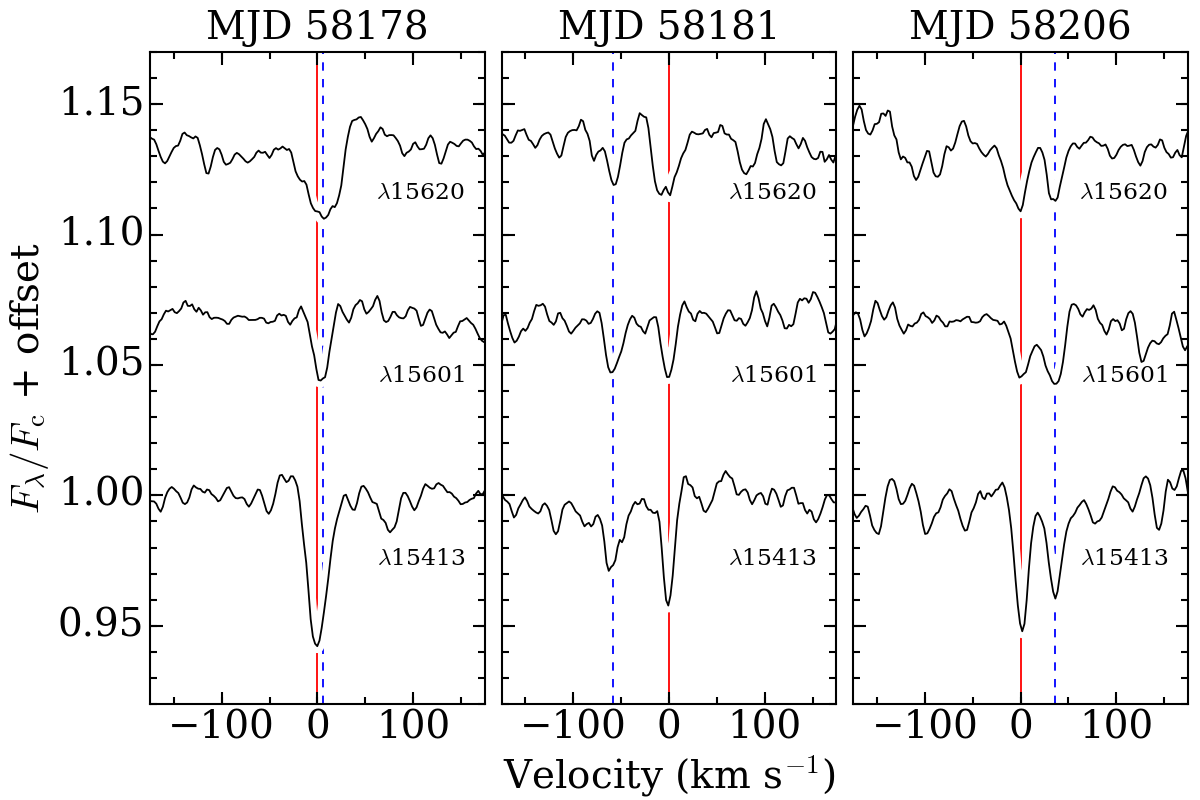}
\caption{Multiple sets of {\mnii} lines in the $H$-band spectra of the SB2 2MASS\,J09432220$-$5348264 (HD\,298641). The {\mnii}~15413, 15601, and 15620\,{\AA} line profiles are shown in the rest frame of the star with stronger {\mnii}, with a vertical dashed line indicating the position of the secondary star lines. }
\label{fig09432220}
\end{figure}

\begin{figure}
\centering 
\includegraphics[width=1.0\columnwidth]{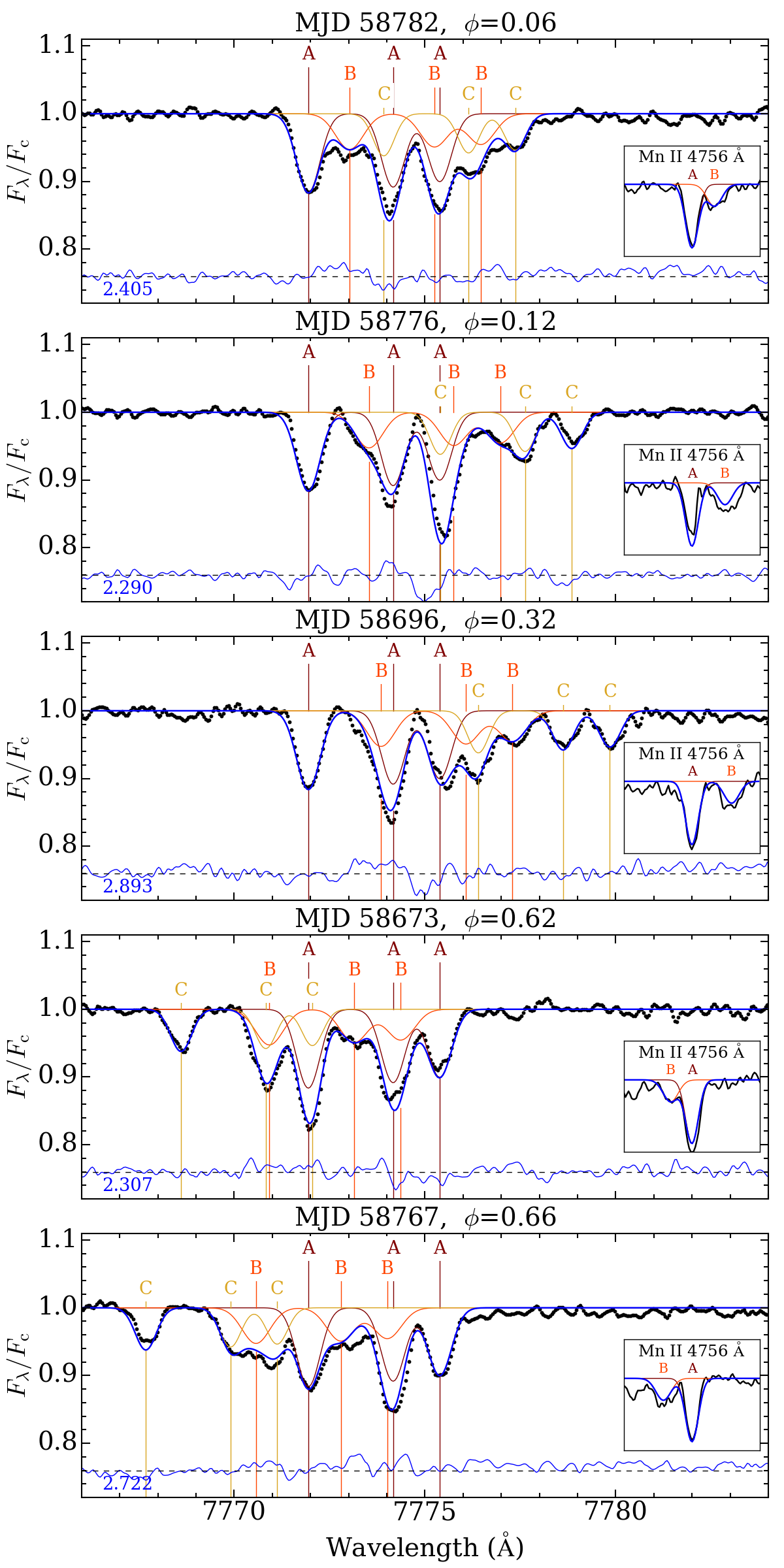}
\caption{Optical spectra of the HgMn+Bp\,Mn+Am SB3 2MASS\,J19223868+1433117 (HD\,231263), with the large panels showing the {\oi} triplet (7772, 7774, 7775\,{\AA}) blend to which all three stars contribute and with the smaller panels showing an example of the {\mnii} lines that are produced by the HgMn (A) and Bp\,Mn (B) stars but not by the Am (C) star. Meanings are otherwise the same as in Figure~\ref{fig00570626}.}
\label{fig19223868}
\end{figure}

\noindent\textbf{2MASS\,J02060290+5009136} is an SB2 with an Ap/Bp companion that is detected in {\ceiii}~15961\,{\AA} in three of the four available APOGEE spectra. The stronger {\ceiii}~16133\,{\AA} line is unfortunately blended with residuals from the strong airglow line at 16128\,{\AA}, such that we could only measure RVs and estimate {\vsini} based on {\ceiii}~15961\,{\AA}. The spectra also only cover a short timescale of 21 days, but nonetheless, the {\ceiii} RVs are clearly anti-correlated with those of the {\mnii} lines, with {\ceiii} going from $-41$\,{\kms} to $-21$\,{\kms} and with {\mnii} going from $-34$\,{\kms} to $-42$\,{\kms}. More spectra will be needed to determine the orbital parameters of the system and to confirm that is indeed another example of the rare HgMn+Ap/Bp binaries.

\vspace{0.1in}

\noindent\textbf{2MASS\,J09432220$-$5348264 = HD\,298641} is the only available example of an HgMn+HgMn binary in which multiple sets of {\mnii} lines were detected in the APOGEE spectra. Figure~\ref{fig09432220} shows the {\mnii}~15413, 15600, and 15620\,{\AA} lines from the three spectra, with vertical lines denoting the lines from each star. The components were fully blended on MJD 58178 but well separated on MJDs 58181 and MJD 58206. Although the spectra in Figure~\ref{fig09432220} are shown in the rest frame of the star that produces stronger {\mnii}~15413\,{\AA}, it is in fact the slightly more RV-variable star (RV$_{\rm scat}$ of 48.6\,{\kms} versus 46.2\,{\kms}).  Based on the available data, it appears likely that the orbital period is less than 50 days.

\vspace{0.1in}

\noindent\textbf{2MASS\,J18064814$-$2902119 = HD\,165347} is a known visual double star with a far fainter companion (5 magnitudes fainter) separated by 6.85{\arcsec} \citep{mason2001}. The 11 APOGEE spectra of the system instead indicate that it is at least an SB3. One of the companions is an Ap/Bp star whose {\ceiii} lines and $UL$ ({\ceiii}~15961\,{\AA} being the strongest line) are detected in 9/11 spectra, and the other is an A/Am star whose $NSL$ ({\sili}~15964\,{\AA} being the strongest line) are detected in 7/11 spectra. The dominant component of the spectra is of course the HgMn star, but over the 6.2 year observational baseline, the {\mnii} RVs vary by just 10\,{\kms} with an average of $-35$\,{\kms}. The Ap/Bp and A/Am stars are highly RV-variable however, with RV$_{\rm scat}$ of 130 and 133\,{\kms}, respectively. Figure~\ref{sb2orbits} shows an SB2 orbital solution that suggests they are on a slightly eccentric orbit with $P$=5.47 days and that the A/Am star is slightly more massive than the Ap/Bp star. The fact that the systemic velocity of the tight SB2 differs from the average RV of the HgMn star by almost 18\,{\kms} may imply that the HgMn star is on a very long period orbit around either the inner SB2 or a fourth object. 

\vspace{0.1in}

\noindent\textbf{2MASS\,J19223868+1433117 = HD\,231263} is one of the more remarkable discoveries presented here. It is a known visual double star with a companion of comparable brightness ($\sim0.2$ magnitude difference) that is separated by just 0.5{\arcsec} \citep{mason2001}. This may be the Am SB2 companion that is detected in {\ci}~16895\,{\AA} and the {\mgi} lines (15753, 15770\,{\AA}) in the six APOGEE spectra, but our five optical spectra of the system show that it is actually a multiple system composed of at least three stars. The third star apparently does not produce any strong lines in the $H$-band.

All three stars in the HD\,231263 system produce {\mgii}, {\silii}, {\tiii}, {\feii}, {\oi}, and Balmer series lines in the optical, and detection of multiple sets of {\hei} lines indicates that the secondary (undetected in the $H$-band) is a late-B type star. Simultaneous detection of {\mnii} lines (though not {\hgii}) at the right RVs further indicates that this late-B companion is in fact also an Mn-peculiar star, such that this is an HgMn+Bp\,Mn+Am system. However, it was the cooler tertiary that was detected in the APOGEE spectra and that is on a circular orbit around the HgMn star. Figure~\ref{sb2orbits} shows the associated SB2 orbital solution, which has a relatively short period of just $\sim$6.2 days. 

Figure~\ref{fig19223868} shows an attempt to de-blend the {\oi} triplet lines using three sets of Gaussians with fixed equivalent width ratios based on the expected relative strengths of the lines. For the HgMn star, $W(7772$\,{\AA}) was fixed to the value measured from the apparently unblended line on MJD 58696, while for the Am star (`C'), $W(7772$\,{\AA}) was fixed to the average value measured from the apparently unblended lines on MJDs 58673 and 58767. For the Bp\,Mn star (`B'), the {\oi} wavelengths and widths were fixed based on measurements of {\feii}, {\silii}, and {\mgii} lines blueward of H$\alpha$, and the equivalent widths were simply adjusted until a satisfactory fit of the full blend was achieved. This resulted in $W(7772$\,{\AA}) of 90, 48, and 43\,m{\AA} for the A, B and C components, respectively.

The RVs of the Bp\,Mn star vary by just 4.7\,{\kms} over the 109 days covered by the optical spectra, but the average RV of $-7.3$\,{\kms} is quite close to the $-17.1$\,{\kms} systemic velocity of the HgMn+Am SB2. It may be the case that the Bp\,Mn star is on a long period orbit around a fourth star, but it is certainly associated with the SB2.

\subsection{Binaries with Am Companions}
\noindent\textbf{2MASS\,J02261624+5704069 = HD\,14900} is a known visual double star, with a far fainter companion (4.1 magnitudes fainter) separated by 13{\arcsec} \citep{mason2001}. Although we initially classified this system as an SB1, having found a short orbital period of 4.5 days based on the 10 APOGEE spectra, the companion was clearly detected in three subsequent optical spectra. The strongest line for the companion is {\mgii}~4481\,{\AA} ($W_{\lambda}=24$\,m{\AA}), but the \ion{Ba}{ii} resonance line at 4554\,{\AA} is of comparable strength ($W_{\lambda}=14$\,m{\AA}), suggesting that the companion is probably an Am star. Both stars are extremely slow rotators with {\vsini}\,$\sim$\,3--4\,{\kms}. The orbital solution shown in Figure~\ref{sb2orbits} indicates a mass ratio of $q=0.542$, which is among the smallest of the SB2 systems discussed in this paper. This might explain the $H$-band non-detection of the companion.

\vspace{0.1in}

\noindent\textbf{2MASS\,J04502564+6328354} is an SB2 with an Am companion that is clearly detected in the {\mgi} lines and in {\ci}~16895\,{\AA} in the nine available APOGEE spectra and in numerous lines in two optical spectra. An orbital solution of the system is shown in Figure~\ref{sb2orbits}, indicating a short period of just 4.3 days and a mass ratio of 0.687. The lack of {\hei} lines at the companion velocities in the two optical spectra confirms the associated A spectral type, and this is further supported by the companion's low-energy {\mgi} and {\fei} lines that lack counterparts from the HgMn star. The SB2 components were well-separated on one of the optical epochs (see Figure~\ref{specmontage2}), allowing for unambiguous identification of number {\crii}, {\srii}, and \ion{Ba}{ii} lines from the companion that are quite strong compared to {\mgii}~4481\,{\AA}. For example, the flux ratio of {\srii}~4078\,{\AA}/{\mgii}~4481\,{\AA} is about 0.4, such that the companion is almost certainly an Am star.

\vspace{0.1in}

\noindent\textbf{2MASS\,J05380741+2418520 = HD\,37242} is an SB2 in which $NSL$ from the Am companion were detected in just one of three APOGEE spectra, with the $NSL$ being offset from the {\mnii} lines by $\sim$166\,{\kms} on that epoch. The system was subsequently confirmed as an SB2 by an optical spectrum in which the companion contributions are quite obvious. As with the other SB2s discussed in this section, we classify the companion as an Am star on the basis of overly strong {\srii} and {\baii} lines. In this case, the equivalent width of the companion's {\baii}~4554\,{\AA} line is nearly half that of {\mgii}~4481\,{\AA}. Unlike most of the SB2 systems discussed here, the companion ({\vsini}\,=\,15\,{\kms}) is a far slower rotator than the HgMn star ({\vsini}\,=\,58\,{\kms}). More spectra will be needed for a full orbital analysis, but based on the large velocity separations of the SB2 components during the two epochs where both were detected, the orbital period is likely to be very short. 

\vspace{0.1in}

\noindent\textbf{2MASS\,J07144161$-$1017068} is SB2 with an Am companion whose $NSL$ were detected in only two of the five APOGEE spectra. The system is relatively faint ($H=10.366$) compared to most of the sample, such that the companion's weak lines are only discernible after considerable smoothing of the spectra. Our optical follow-up spectrum confirmed that the system is indeed an SB2, and also that the companion is probably an Am star based on clear detection of the \ion{Ba}{ii} lines. Both of the stars are extremely slow rotators, with nearly identical {\vsini} of $\sim$6\,{\kms}. Despite only 6 total spectra, the orbital period was constrained to $<10$ days by the APOGEE spectra (separated by 17 days) in which the companion was detected. An orbital solution with $e=0$ and $P=9.25$ days is shown in Figure~\ref{sb2orbits}.

\vspace{0.1in}

\noindent\textbf{2MASS\,J17314436$-$2616112 = HD\,158704} is a previously known HgMn star identified as an SB2 by \citet{dolk2003}. It is a visual double star with a 0.35{\arcsec} separation \citet{mason1999} and a clear SB2 in both the optical and $H$-band. A third companion was identified in the high-contrast images presented by \citet{scholler2010}, and based on the relatively constant RVs (RV$_{\rm scat}$ of 5.7\,{\kms} across 2 APOGEE spectra and 3 optical spectra) of of the HgMn star, we suspect that this is indeed a multiple system. The slowly-rotating Am companion is certainly RV-variable however, with RV$_{\rm scat}=37.8$\,{\kms}.

\vspace{0.1in}

\noindent\textbf{2MASS\,J20360994+3929487} is one of the more obvious SB2s of the sample, with numerous narrow $NSL$ detected in the 6 APOGEE spectra. The companion is a slowly-rotating star with {\vsini}\,$\sim11$\,{\kms}, and considering the relative weakness of its lines in the optical, it may be the coolest SB2 secondary of the sample. For example, attempts to deblend the contributions to {\mgii}~4481\,{\AA} from the two stars indicates that the companion line is perhaps only 10\% as strong the HgMn star line. Nonetheless, {\baii} lines from the companion are clearly present in the spectra at strengths comparable to {\mgii}~4481\,{\AA}, such that we classify it as an Am star. Figure~\ref{sb2orbits} shows an orbital solution with a 31.4 day period and a relatively high eccentricity of $e=0.58$. The mass ratio of $M_{2}/M_{1}=0.536$ is the lowest reported in this paper.  

\vspace{0.1in}

\noindent\textbf{2MASS\,J23575323+5723194 = HD\,224435} is another known visual multiple system, with at least two companions separated by 1.2{\arcsec} and 8.7{\arcsec}, respectively \citep{mason2001}. The more distant companion is about 3.8 magnitudes fainter than the HgMn star, but the close companion is only 0.2 magnitudes fainter and hence is presumably the SB2 companion detected in the 18 APOGEE spectra and 4 optical spectra. The companion is a slow rotator with {\vsini}\,$\sim11$\,{\kms} and with the {\mgi} and {\ci} lines being the strongest $H$-band features. As with the other stars discussed in this section, we classify the companion as Am based on clear detection of {\srii} and {\baii} lines that are comparable in strength to {\mgii}~4481\,{\AA}. Despite the Am star being highly RV-variable, the RVs of the HgMn star vary by only $\sim13$\,{\kms} over the 8.2 year observational baseline, with variation pattern seeming unrelated to the motion of the Am star. Figure~\ref{sb2orbits} shows an SB1 orbital solution for the Am star, with the HgMn star RVs phased to the 21 day period. The average RV of the HgMn star is quite close to the systemic velocity indicated by the SB1 orbit of the Am star (-10.7\,{\kms} versus -7.2\,{\kms}), such that the two stars are certainly associated. 

\subsection{Other SB2s and SB3s}

\noindent\textbf{2MASS\,J02203397+6023494} is a likely multiple system. The spectrum of an A/Am star was detected in all 11 APOGEE spectra, with the RVs varying by 11\,{\kms}. The HgMn star RVs are even more highly RV-variable, with the maximum difference between epochs being 18\,{\kms}. Despite this, we could not find reasonable orbital solutions for either star and certainly not for an SB2 combination since the relative RV variability seems random.

\vspace{0.1in}

\noindent\textbf{2MASS\,J04315476+2138086 = HD\,28662} is a definite SB2 with an A or Am companion that is the fastest rotator ({\vsini}\,$\sim$\,62\,{\kms}) of all the $H$-band-detected secondary stars described in this paper. In the single APOGEE spectrum of the system, the companion is most clearly detected in the {\mgi} lines between 15745--15770\,{\AA} and in {\ci}~16895\,{\AA}, with those lines being offset from {\mnii} by roughly 149\,{\kms}.

\vspace{0.1in}

\noindent\textbf{2MASS\,J04510073+3226501 = HD\,30661} is another SB2 in which the companion was detected in optical spectra but not in the APOGEE spectra. An SB2 orbital solution shown in Figure~\ref{sb2orbits}, and the associated mass ratio of $q=0.895$ is the largest reported here. The optical data show that the stars are near spectroscopic twins in terms of almost identical {\vsini} and almost identical depths and strengths of the Balmer series, {\silii}, {\feii}, {\tiii}, {\mgii}, and {\oi} lines (among other species). The two most striking differences between the two stars are the lack of {\mnii}, {\hgii}, and other heavy metal lines from the companion star, as well as the relative strength of the {\hei} lines for the two stars. Namely, the {\hei}~5876\,{\AA} equivalent width of the HgMn star is about half that of the companion. This would normally suggest that the companion is in fact the hotter, more massive star, but in this case it is almost certainly due to a severe depletion of He for the HgMn star. 

\vspace{0.1in}

\noindent\textbf{2MASS\,J06402199$-$0322092 = HD\,47798} is a visual double star confirmed by \citet{mason1999}. We measured RVs from possible weak $NSL$ features in the three APOGEE spectra covering 57 days, indicating that there may be an A/Am companion. However, the resulting RVs are a good match to those of the HgMn star, with the averages being 22\,{\kms} and 23\,{\kms}, respectively. The RV$_{\rm scat}$ is also small for both, at 5.5\,{\kms} for the HgMn star and 2.5\,{\kms} for the A/Am star. This may be a long-period SB2 or a multiple system.

\vspace{0.1in}

\noindent\textbf{2MASS\,J06561603$-$2839053 = HD\,51459} is another SB2 with an A/Am companion. In the single APOGEE observation, the RVs of the HgMn and A/Am components were 65\,{\kms} and $-79$\,{\kms}, respectively. Relatively strong {\ceiii} lines (e.g, {\ceiii}~15961\,{\AA}/{\mnii}~15413\,{\AA}\,$\sim0.53$) are present in the spectra, but their RVs and FWHM are all but identical to those of the {\mnii} lines. 

\vspace{0.1in}

\noindent\textbf{2MASS\,J06572439$-$0928136 = HD\,51539} is another SB2 with an A/Am companion that is clearly detected in the 3/4 available APOGEE spectra with sufficient S/N. The RVs of the HgMn and A/Am star vary by just 4.4\,{\kms} and 2.5\,{\kms}, respectively, in the three spectra taken over 50 days, but the variability is anti-correlated. 

\vspace{0.1in}

\noindent\textbf{2MASS\,J14482274$-$5959316 = HD\,130039} was included in the sample due to fairly confident detection of {\mnii}~15413 and 15620\,{\AA} in the first of three APOGEE spectra (MJD 58657). That spectrum also exhibits numerous $NSL$ (indicating an A/Am spectral type) at an RV of 98\,{\kms} as compared to the {\mnii} RV of 0\,{\kms}. The {\mnii} lines are not confidently detected in the subsequent two spectra (MJDs 56677 and 56689), but the A/Am star RVs drop to $-23$\,{\kms} and then $-133$\,{\kms} in those spectra. This is certainly an SB system, but it is unclear whether it actually involves an HgMn star. It is a known visual double star with a small separation of 0.56{\arcsec}, but the visual companion is 4.24 magnitudes fainter than the primary star \citep{horch2015}.

\vspace{0.1in}

\noindent\textbf{2MASS\,J16133258$-$6019519} is a likely SB2 with an A/Am companion. The {\mgi} lines at 15753 and 15770\,{\AA} seem to definitely be present in the spectra, but disagreement of the RVs of the lines renders the result ambiguous. The HgMn star is only marginally RV-variable.

\vspace{0.1in}

\noindent\textbf{2MASS\,J18391486+2029512 = HD\,172397} has a late-B SB2 companion that was detected in our five optical follow-up spectra but not in the seven APOGEE spectra. The $H$-band non-detection is almost certainly due in part to the relatively rapid rotation of the companion; based on fitting multiple Gaussians to the partially blended {\silii}~3863\,{\AA}, {\mgii}~4481\,{\AA}, and {\feii}~5169\,{\AA} lines in one of the optical spectra, we estimate that the companion has {\vsini}$\sim70$\,{\kms}. Any $H$-band metallic features from the star would be quite weak to begin with, and probably undetectable for such a rapid rotator. The ratio of the equivalents of {\hei}~4471\,{\AA} over {\mgii}~4481\,{\AA} for the companion is not much less than unity, indicating that it is definitely hotter than A0. The lack of detectable metal absorption features from the companion in the $H$-band is therefore unsurprising (see Figure~\ref{specmontage}). Figure~\ref{sb2orbits} shows a possible SB2 orbit with a 35 day period and $e=0.36$. 

\vspace{0.1in}

\noindent\textbf{2MASS\,J19294384$-$1801510 = HD\,183327} has an A/Am companion whose $NSL$ are quite obvious and offset from the {\mnii} lines by $\sim13$\,{\kms} in the single APOGEE spectrum. 

\vspace{0.1in}

\noindent\textbf{2MASS\,J19454054+3959248 = HD\,225692} has a broad-lined companion that was detected in our three optical spectra but not in the 13 APOGEE spectra. The broad components are most obvious in the {\hei} lines and {\mgii}~4481\,{\AA}, with weaker contributions in the {\silii} and {\oi} lines, indicating that it is a late-B type star. Based on attempts to fit Gaussians to the blended {\hei}~4471\,{\AA} and {\mgii}~4481\,{\AA} region, we estimate that the companion has {\vsini}$\sim200$\,{\kms}. Thus, it is not terribly surprising that we see no evidence of this star in the APOGEE spectra, since it might not produce significant metal absorption lines even if was not such a rapid rotator. The orbital solution shown in Figure~\ref{sb2orbits} indicates $P=338$ days, which matches the period found if we treat the system as an SB1, and which is by far the longest orbital period reported in this paper. The mass ratio of the SB2 orbital solution ($M_{2}/M_{1}=0.538$) is among the lowest reported in this paper, which is surprising considering that the companion is a B-type star. 

\section{Carbon Emission} \label{c1em}
As mentioned in Section~\ref{sample} and Figure~\ref{specmontage}, HgMn stars often exhibit conspicuous narrow emission in the {\ci}~16895\,{\AA} line. More specifically, this is true for roughly 45\% of the sample (120 stars). Only in cases of exceptionally strong {\ci}~16895\,{\AA} emission are the weaker {\ci} lines at 16009 and 16026\,{\AA} (labeled in Figure~\ref{specmontage}) observed in emission. To confirm that the emission is in fact associated with the HgMn stars rather than being some kind of spurious artifact, we fit Gaussians to the {\ci} emission features in order to obtain estimates of the RV, FWHM, and equivalent width of the feature. 

\begin{figure}
\includegraphics[width=1.0\columnwidth]{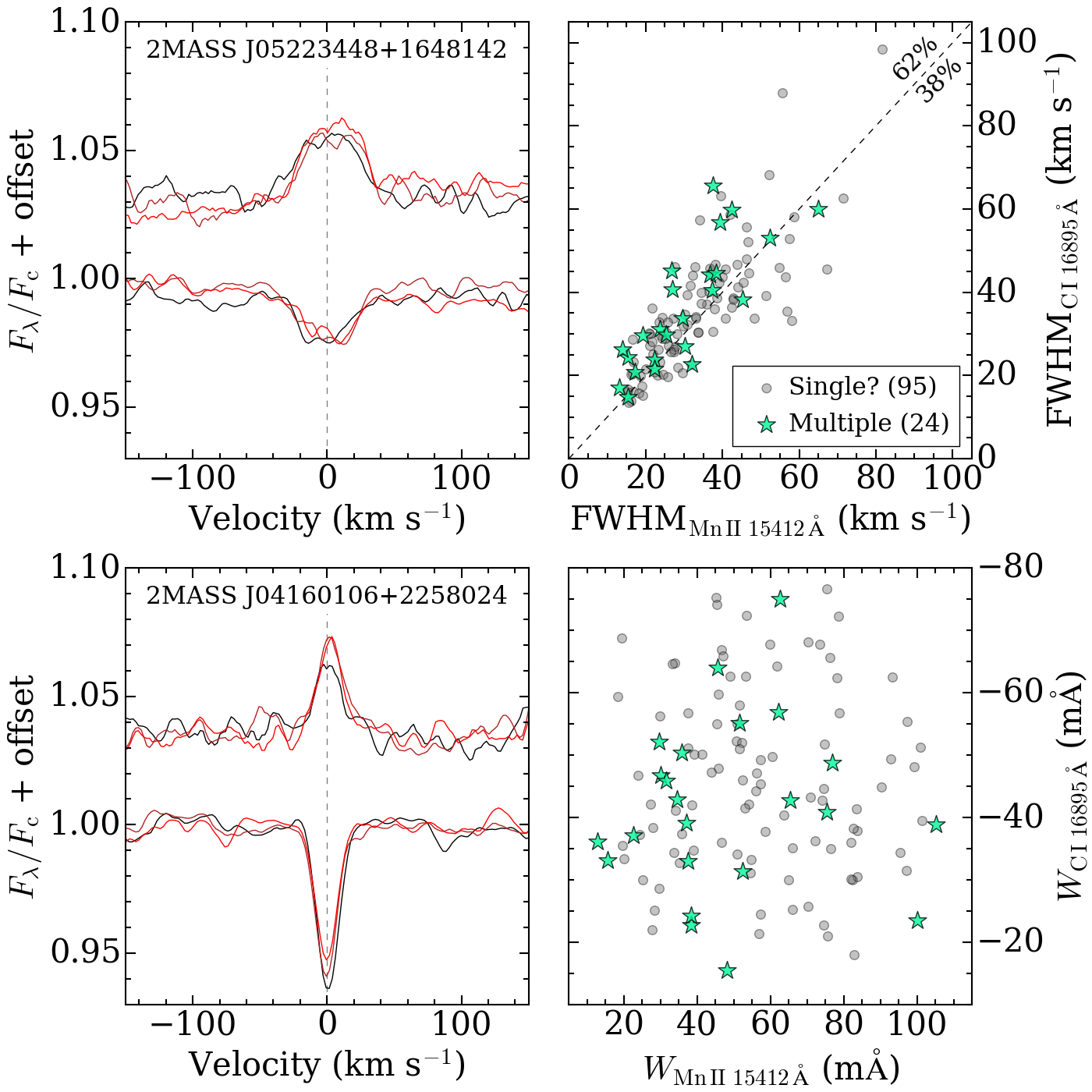}
\caption{(\emph{Left panels:}) Demonstration of the matching radial velocities and line widths of {\mnii}~15413\,{\AA} absorption and {\ci}~16895\,{\AA} emission. Each panel shows spectra from three epochs, corrected to rest frame based on the {\mnii} radial velocities and with the {\ci} region artificially shifted upward. 2MASS designations are given above each panel. (\emph{Upper right:}) The FWHMs of {\mnii}~15413\,{\AA} versus those of {\ci}~16895\,{\AA}, with the dotted line indicating $y=x$. (\emph{Lower right:}) The equivalent widths of the same lines. Star symbols indicate binary and multiple systems.}
\label{proveC1}
\end{figure}

Based on this, our first conclusion regarding the nature of the line emission is demonstrated in the two lefthand panels of Figure~\ref{proveC1}. These panels show the {\mnii}~15413\,{\AA} (lower lines) and {\ci}~16895\,{\AA} (upper lines) line profiles from three epochs for two stars that exemplify the behavior of these lines. The spectra have been shifted to rest frame based on the RVs of the {\mnii} lines, making it clear that the {\ci} emission occurs at roughly the same RVs is therefore definitely associated with the HgMn stars. However, small blueshifts (1--5\,{\kms}) of the {\ci} emission peaks are visible in most of the displayed spectra, indicating that the emission forms just above the surfaces of the stars. 

\begin{figure}
\includegraphics[width=1.0\columnwidth]{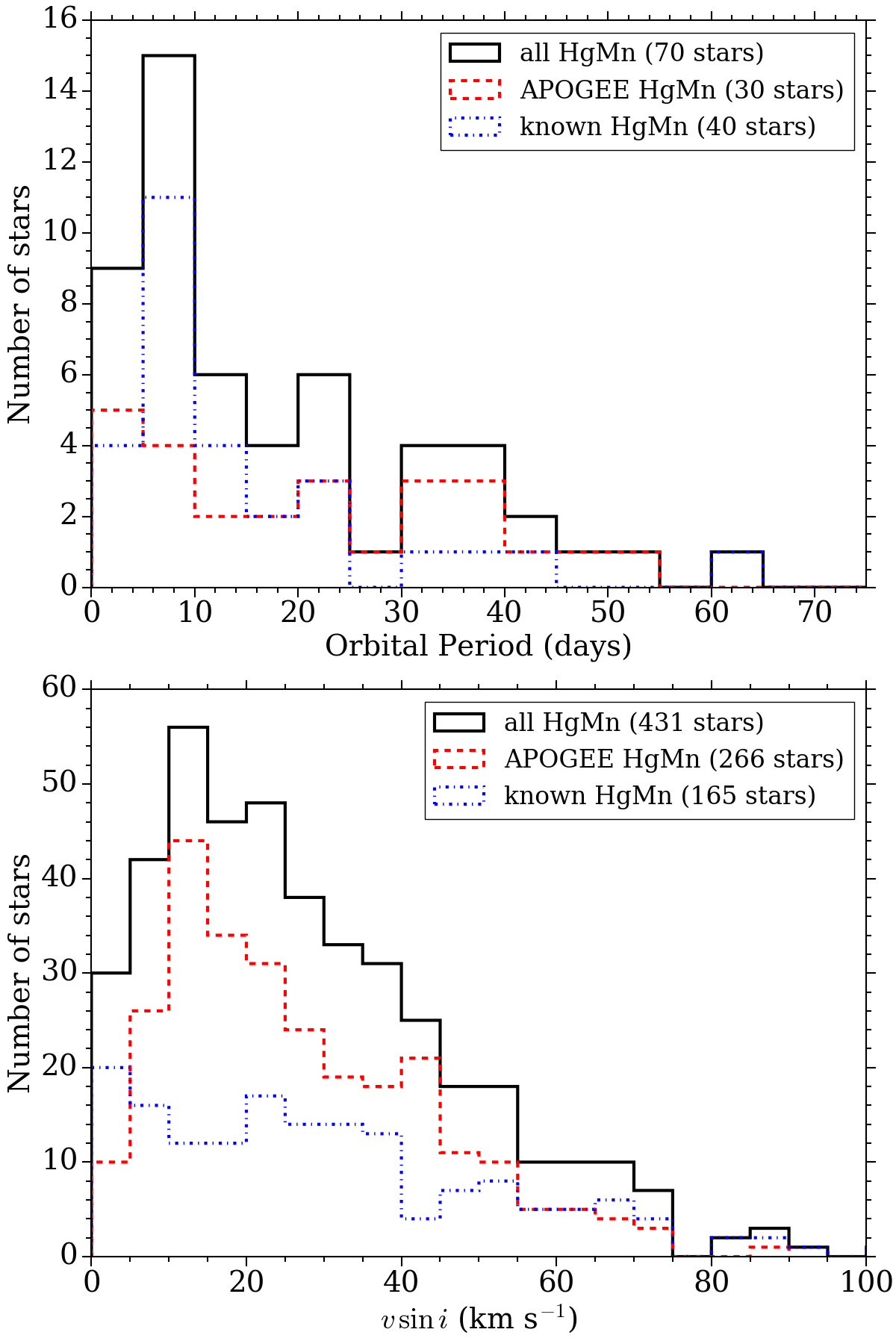}
\caption{Histogram of orbital periods for HgMn binaries. The solid line indicates the APOGEE sample, and the dashed line indicates the known sample.}
\label{periodHist}
\end{figure}

Our second conclusion is the one made obvious by inspection of the line profile widths in lefthand panels of Figure~\ref{proveC1}, and then quantified in the upper right panel, which plots the FWHM of the {\ci}~16895\,{\AA} emission versus the FWHM of {\mnii}~15413\,{\AA}. There is considerable scatter about the $y=x$ line, with 62\% of the stars having {\ci} wider than {\mnii}, but the suggestion is nonetheless that the {\ci} emission exhibits the same level of broadening as the {\mnii} lines and is therefore definitely associated with the surfaces of the stars. The same conclusions were reached by \citet{wahlgren2000} with respect to narrow emission lines in the optical spectra of chemically normal and chemically peculiar late-B stars.

As far as correlations with other spectral features, there are none to be found in the APOGEE spectra. The lower right panel of Figure~\ref{proveC1} plots the equivalents widths of {\mnii}~15413\,{\AA} versus those of {\ci}~16895\,{\AA}, showing that the strength of the {\ci} emission has no dependence on the strength of the {\mnii} absorption lines. There also does not seem to be a correlation between {\ci} emission and binarity, since the fraction of stars with {\ci} emission that are definitely binaries or multiples (20\%) is considerably lower than the 31\% binary fraction for the full sample.

Emission in the {\ci}~16895\,{\AA} line is certainly not a rare phenomenon and is not limited to chemically peculiar stars; recall that the B8\,V star HD\,2559 exhibits {\ci} emission in the spectrum shown in Figure~\ref{specmontage}. {\ci} emission is also present in $\sim20$\% of the classical Be star spectra obtained by APOGEE \citep{choj2015}, and the emission is particularly strong in the case of exotic binaries such as Be+sdO \citep[HD\,55606;][]{choj2018} and Be+black hole \citep[HD\,215227;][]{casares2014}. The {\ci}~16895\,{\AA} line is also observed in emission in the spectra of blue supergiants (e.g., HD\,15266), pre-main-sequence, supergiant, and unclassified B[e] stars (e.g., MWC\,361, MWC\,137, and MWC\,922), luminous blue variables (e.g., S\,Doradus), post-AGB stars (e.g., IRAS\,05040+4820), and extreme Helium stars (e.g., V2244\,Oph). 

\section{Discussion}\label{sect:disc}
Thanks to the strategy of selecting as telluric standards thousands of blue stars across the sky down to $H=11$, the APOGEE survey has more than doubled the known sample of HgMn stars, expanding it from $\sim194$ stars to a grand total of just over 450 stars. Although still quite rare, we estimate that somewhere between 5--10\% of the late-B stars observed by APOGEE are in fact HgMn stars. This significant increase of the sample of HgMn stars promises to open new perspectives for our knowledge and understanding of mid to late-main sequence B-type stars, and some inferences can already be made based on the basic analysis presented in this paper.

Combining the 30 orbital solutions for new HgMn stars with the 40 orbital solutions for known HgMn stars in the 9th Catalogue of Spectroscopic Binary Orbits \citep{pourbaix2009}, we have orbital period estimates for a total of 70 HgMn binaries. Of these systems, 57/70 (77\%) have periods less than 75 days, 54/70 (74\%) have periods less than 50 days, and 40/70 (57\%) have periods less than 25 days. This result is emphasized in the upper panel of Figure~\ref{periodHist}, which shows that the distribution of orbital periods (excluding the 16 systems with $P>75$ days) is strongly peaked in the 5--10 day range. The previously suggested preference for short orbital periods \citep{hubrig1995} is therefore confirmed. 

The lower panel of Figure~\ref{periodHist} shows the distribution of {\vsini} for HgMn stars, confirming that HgMn stars are slow rotators as a rule. The overall average is $<${\vsini}$>=28$\,{\kms} and the distribution is strongly peaked toward {\vsini}\,$\sim$\,20\,{\kms}. For comparison, the average {\vsini} of non-supergiant, B7--A0, apparently chemically normal stars included in the catalog of \citet{abt2002} is $>120$\,{\kms}. The phenomenon of B-type stars exhibiting HgMn abundance anomalies is therefore intimately linked not only with their multiplicity but with their slow rotation. These are key ingredients in the formation of HgMn stars.

\section*{Acknowledgements}

We dedicate this paper to the memory of Dr. Fiorella Castelli, whose contributions to the study of chemically peculiar stars will not be forgotten. We also thank the anonymous referee for useful comments that improved this paper.

Funding for the Sloan Digital Sky Survey IV has been provided by the Alfred P. Sloan Foundation, the U.S. Department of Energy Office of Science, and the Participating Institutions. SDSS acknowledges support and resources from the Center for High-Performance Computing at the University of Utah. The SDSS web site is \href{www.sdss.org}{www.sdss.org}.

SDSS is managed by the Astrophysical Research Consortium for the Participating Institutions of the SDSS Collaboration including the Brazilian Participation Group, the Carnegie Institution for Science, Carnegie Mellon University, the Chilean Participation Group, the French Participation Group, Harvard-Smithsonian Center for Astrophysics, Instituto de Astrof\'isica de Canarias, The Johns Hopkins University, Kavli Institute for the Physics and Mathematics of the Universe (IPMU) / University of Tokyo, Lawrence Berkeley National Laboratory, Leibniz Institut fur Astrophysik Potsdam (AIP), Max-Planck-Institut fur Astronomie (MPIA Heidelberg), Max-Planck-Institut für Astrophysik (MPA Garching), Max-Planck-Institut für Extraterrestrische Physik (MPE), National Astronomical Observatories of China, New Mexico State University, New York University, University of Notre Dame, Observatório Nacional / MCTI, The Ohio State University, Pennsylvania State University, Shanghai Astronomical Observatory, United Kingdom Participation Group, Universidad Nacional Aut\'onoma de M\'exico, University of Arizona, University of Colorado Boulder, University of Oxford, University of Portsmouth, University of Utah, University of Virginia, University of Washington, University of Wisconsin, Vanderbilt University, and Yale University.

S.H. is supported by an NSF Astronomy and Astrophysics Postdoctoral Fellowship under award AST-1801940. D.A.G.H. acknowledges support from the State Research Agency (AEI) of the Spanish Ministry of Science, Innovation and Universities (MCIU) and the European Regional Development Fund (FEDER) under grant AYA2017-88254-P.




\bsp	
\label{lastpage}
\end{document}